\documentclass[aps,prd,twocolumn,superscriptaddress,preprintnumbers,nofootinbib,longbibliography,floatfix]{revtex4-1}

\usepackage{graphicx}
\usepackage{dcolumn}
\usepackage{bm}
\usepackage{enumitem}

\input{Qcircuit}
\usepackage[colorlinks=true
,urlcolor=blue
,anchorcolor=blue
,citecolor=blue
,filecolor=blue
,linkcolor=blue
,menucolor=blue
,pagecolor=blue
]{hyperref}
\usepackage[all]{hypcap}
\usepackage{braket}
\usepackage{amsmath,amsfonts}
\usepackage{algorithm, algorithmic, float}
\usepackage{comment}
\usepackage{cleveref}

\DeclareRobustCommand{\Sec}[1]{Sec.~\ref{#1}}
\DeclareRobustCommand{\Secs}[2]{Secs.~\ref{#1} and \ref{#2}}

\DeclareRobustCommand{\Tab}[1]{Table~\ref{#1}}

\DeclareRobustCommand{\Fig}[1]{Fig.~\ref{#1}}

\DeclareRobustCommand{\Eq}[1]{Eq.~(\ref{#1})}
\DeclareRobustCommand{\Eqs}[2]{Eqs.~(\ref{#1}) and (\ref{#2})}
\DeclareRobustCommand{\Ref}[1]{Ref.~\cite{#1}}
\DeclareRobustCommand{\Refs}[1]{Refs.~\cite{#1}}

\DeclareMathOperator*{\argmax}{argmax}
\newcommand{\opt}{\text{opt}}

\def\begsub#1#2\endsub{\begin{subequations}\label{eq:#1}\begin{align}#2\end{align}\end{subequations}}
\def\ba#1\ea{\begin{align}#1\end{align}}
\def\bas#1\eas{\begin{align*}#1\end{align*}}

\usepackage{acronym}

\begin{document}

%
%

\preprint{MIT-CTP 5137}

\title{Quantum Algorithms for Jet Clustering}

\author{Annie Y. Wei}
\email{anniewei@mit.edu}
\affiliation{Center for Theoretical Physics, Massachusetts Institute of Technology, Cambridge, MA 02139}

\author{Preksha Naik}
\email{prekshan@mit.edu}
\affiliation{Center for Theoretical Physics, Massachusetts Institute of Technology, Cambridge, MA 02139}

\author{Aram W. Harrow}
\email{aram@mit.edu}
\affiliation{Center for Theoretical Physics, Massachusetts Institute of Technology, Cambridge, MA 02139}

\author{Jesse Thaler}
\email{jthaler@mit.edu}
\affiliation{Center for Theoretical Physics, Massachusetts Institute of Technology, Cambridge, MA 02139}
\affiliation{Department of Physics, Harvard University, Cambridge, MA 02138}

\begin{abstract}
Identifying jets formed in high-energy particle collisions requires solving optimization problems over potentially large numbers of final-state particles.
In this work, we consider the possibility of using quantum computers to speed up jet clustering algorithms.
Focusing on the case of electron-positron collisions, we consider a well-known event shape called thrust whose optimum corresponds to the most jet-like separating plane among a set of particles, thereby defining two hemisphere jets.
We show how to formulate thrust both as a quantum annealing problem and as a Grover search problem.
A key component of our analysis is the consideration of realistic models for interfacing classical data with a quantum algorithm.
With a sequential computing model, we show how to speed up the well-known $O(N^3)$ classical algorithm to an $O(N^2)$ quantum algorithm, including the $O(N)$ overhead of loading classical data from $N$ final-state particles.
Along the way, we also identify a way to speed up the classical algorithm to $O(N^2 \log N)$ using a sorting strategy inspired by the \textsc{SISCone} jet algorithm, which has no natural quantum counterpart.
With a parallel computing model, we achieve $O(N \log N)$ scaling in both the classical and quantum cases.
Finally, we consider the generalization of these quantum methods to other jet algorithms more closely related to those used for proton-proton collisions at the Large Hadron Collider.
\end{abstract}

\maketitle

{\linespread{0.98} \footnotesize \tableofcontents}

%
%

\acrodef{QCD}{quantum chromodynamics}
\acrodef{QUBO}{quadratic unconstrained binary optimization}
\acrodef{LHC}{Large Hadron Collider}

%
%

\section{Introduction}
\label{sec:intro}

Jets are collections of collimated, energetic hadrons formed in high-energy particle collisions.
With an appropriate choice of jet clustering algorithm~\cite{Salam:2009jx}, jets are a robust probe of \ac{QCD} and a useful proxy for determining the kinematics of the underlying hard scattering process.
The problem of identifying jets from collision data is a nontrivial task, however, since the jet clustering algorithm must be matched to the physics question of interest.
Moreover, it is a computationally intensive task, as it often involves performing optimizations over potentially large numbers of final-state particles.

In this paper, we consider the possibility of using quantum computers to speed up jet identification.
We focus on the well-known problem of partitioning an electron-positron collision event into two \emph{hemisphere jets}, though our results are relevant for other optimization problems beyond high-energy physics.
Our main results are summarized in \Tab{tab:summary}, where the computational scaling is given for $N$ particles in the final state.
We show how to improve the well-known $O(N^3)$ classical algorithm~\cite{Yamamoto:1984fd} to an $O(N^2)$ quantum algorithm, which includes the cost of loading the classical data into a sequential quantum computing architecture.
On the other hand, we also show how to speed up the classical algorithm to $O(N^2 \log N)$, using a clever sorting strategy from \Ref{Salam:2007xv}, which matches the quantum performance up to $\log N$ factors.
Finally, using parallel computing architectures, we achieve $O(N \log N)$ scaling in both the classical and quantum cases, albeit for very different computational reasons.

\begin{table*}[t]
\begin{tabular}{l @{$\quad$} l @{$\quad$} l @{$\quad$} l }
\hline
\hline
\textbf{Implementation} & \textbf{Time Usage} & \textbf{Qubit Usage} & \textbf{Section}\\
\hline
\hline
Classical~\cite{Yamamoto:1984fd} & $O(N^3)$ & --- & \Sec{sec:classical_paritioning}\\
Classical with Sort~(using \cite{Salam:2007xv}) & $O(N^2 \log N)$ & --- & \Sec{sec:classical_sort}\\
Classical with Parallel Sort & $O(N \log N)$ & --- & \Sec{subsec:parallel_classical}\\
\hline
Quantum Annealing & Gap Dependent & $O(N)$ & \Sec{sec:qubo}\\
Quantum Search: Sequential Model & $O(N^2)$ & $O(\log N)$  & \Sec{subsec:groverseq}\\
Quantum Search: Parallel Model & $O(N \log N)$ & $O(N\log N)$ & \Sec{subsec:groverpar}\\
\hline
\hline
\end{tabular}
\label{tab:summary}
\caption{Summary of classical and quantum thrust algorithms, where the asymptotic scaling is for a single collision event with $N$ particles.
All strategies have a classical space overhead of $O(N)$ bits for read access to the classical data.
The classical sorting strategies also require write access to $O(N \log N)$ bits.
For ease of exposition throughout, we treat each real number as being specified to a constant $O(1)$ bits of precision.
}
\end{table*}

Quantum algorithms have been shown to achieve speedups over classical algorithms~\cite{Nielsen:2011:QCQ:1972505}, resulting, in theory, in time savings which are even more pronounced over large data sets.
That said, many proposed quantum algorithms for machine learning tasks often omit considerations that would be needed to actually implement them in practice, such as a strategy to interface classical data with a quantum computing architecture.
One solution is to assume the availability of qRAM~\cite{PhysRevLett.100.160501}, which would let our quantum computer access a classical data set in superposition; however this additional hardware requirement may not be easy to implement in practice.
Here, we consider realistic applications of both quantum annealing~\cite{PhysRevE.58.5355, 2000quant.ph..1106F, dwave} and Grover search~\cite{Grover:1996:FQM:237814.237866, Boyer:1996zf, Durr:1996nx} to jet finding, including the $O(N)$ overhead of loading classical collision data into the quantum computer.

The specific jet finding algorithm we use is based on thrust~\cite{Brandt:1964sa,Farhi:1977sg,DeRujula:1978vmq}.
Thrust is an event shape widely measured in electron-positron collisions~\cite{Barber:1979bj,Bartel:1979ut,Althoff:1983ew,Bender:1984fp,Abrams:1989ez,Li:1989sn,Decamp:1990nf,Braunschweig:1990yd,Abe:1994mf,Heister:2003aj,Abdallah:2003xz,Achard:2004sv,Abbiendi:2004qz}.
The optimum value of thrust defines the most jet-like separating plane among a set of final-state particles, thereby partitioning the event into two hemisphere jets.
Algorithmically, it poses an interesting problem because it can be viewed in various equivalent ways---such as a partitioning problem or as an axis-finding problem---which in turn lead to different algorithmic strategies.

We note that practical thrust computations typically involve only 10--1000 particles per event, so the current $O(N^3)$ classical algorithm~\cite{Yamamoto:1984fd} is certainly adequate to the task.
That said, more efficient jet algorithms are of general interest, for example in the context of active area calculations~\cite{Cacciari:2008gn}, which can involve up to millions of ghost particles.
We also note that the current default jet algorithm at the \ac{LHC} is anti-$k_t$~\cite{Cacciari:2008gp}, which already runs in $O(N\log N)$ time~\cite{Cacciari:2005hq,Cacciari:2011ma}, and it is unlikely that any quantum algorithm can yield a sublinear improvement.
On the other hand, anti-$k_t$ is a hierarchical clustering algorithm (i.e.~a heuristic), whereas thrust is a global optimization problem, and there are phenomenological contexts where global jet optimization could potentially yield superior physics performance~\cite{Stewart:2015waa,Thaler:2015xaa}; see also \Refs{Ellis:2001aa,Berger:2002jt,Angelini:2002et,Angelini:2004ac,Grigoriev:2003yc,Grigoriev:2003tn,Chekanov:2005cq,Lai:2008zp,Volobouev:2009rv,Georgi:2014zwa,Ge:2014ova,Bai:2014qca,Mackey:2015hwa,Bai:2015fka}.
Jet finding via global optimization has not seen widespread adoption, in part because of the computational overhead, and we hope the quantum and improved classical algorithms developed here spur more research on alternative jet finding strategies.

Beyond the specific applications to jet finding, we believe that the broader question of identifying realistic quantum algorithms for optimization problems should be of interest to both the particle physics and quantum computing communities.
Indeed, we regard thrust as a warm-up problem for the more general development of quantum algorithms for collider data analysis.
(For other quantum algorithms for collider physics, see \Refs{Mott:2017xdb,Zlokapa:2019lvv} for Higgs boson identification, \Refs{2019arXiv190108148P,Bauer:2019qxa} for parton shower generation, and \Refs{Shapoval:2019txi, Bapst:2019llh,Zlokapa:2019tkn} for track reconstruction.)
Because collider data is classical (and will likely remain so for the foreseeable future), understanding the limitations imposed by data loading is essential to evaluate the potential of quantum algorithms to speed up or improve data analysis pipelines.
At the same time, it is important to assess potential classical improvements to existing collider algorithms, and the sorting strategy of \Ref{Salam:2007xv} is an important example of how new classical strategies can sometimes match the gains from quantum computation.

Turning now to an extended outline of this paper, our quantum algorithms build on existing classical strategies to compute thrust.
In \Sec{sec:def}, we define thrust in its various equivalent manifestations, as both a partitioning problem and an axis-finding problem.
Then in \Sec{sec:classical}, we review classical algorithms for computing thrust based on a search over reference axes.
As already mentioned, the best known result in the literature requires $O(N^3)$ time~\cite{Yamamoto:1984fd}.
We show how to improve it to $O(N^2\log N)$ using a sorting strategy inspired by \textsc{SISCone}~\cite{Salam:2007xv}, which appears to have no quantum analog (see \Sec{subsec:groverseq_sort}).

The first quantum method we consider in \Sec{sec:qubo} involves formulating thrust as a \ac{QUBO} problem, which can then be solved via quantum annealing~\cite{PhysRevE.58.5355, 2000quant.ph..1106F}.
This comes from viewing thrust as a partitioning problem and then considering the brute force enumeration of all candidate partitions.
See \Refs{2018QuIP...17...39K,2018arXiv180302886N} for other studies of quantum annealing for clustering with unique assignment.

\begin{figure*}
\includegraphics[width=\columnwidth]{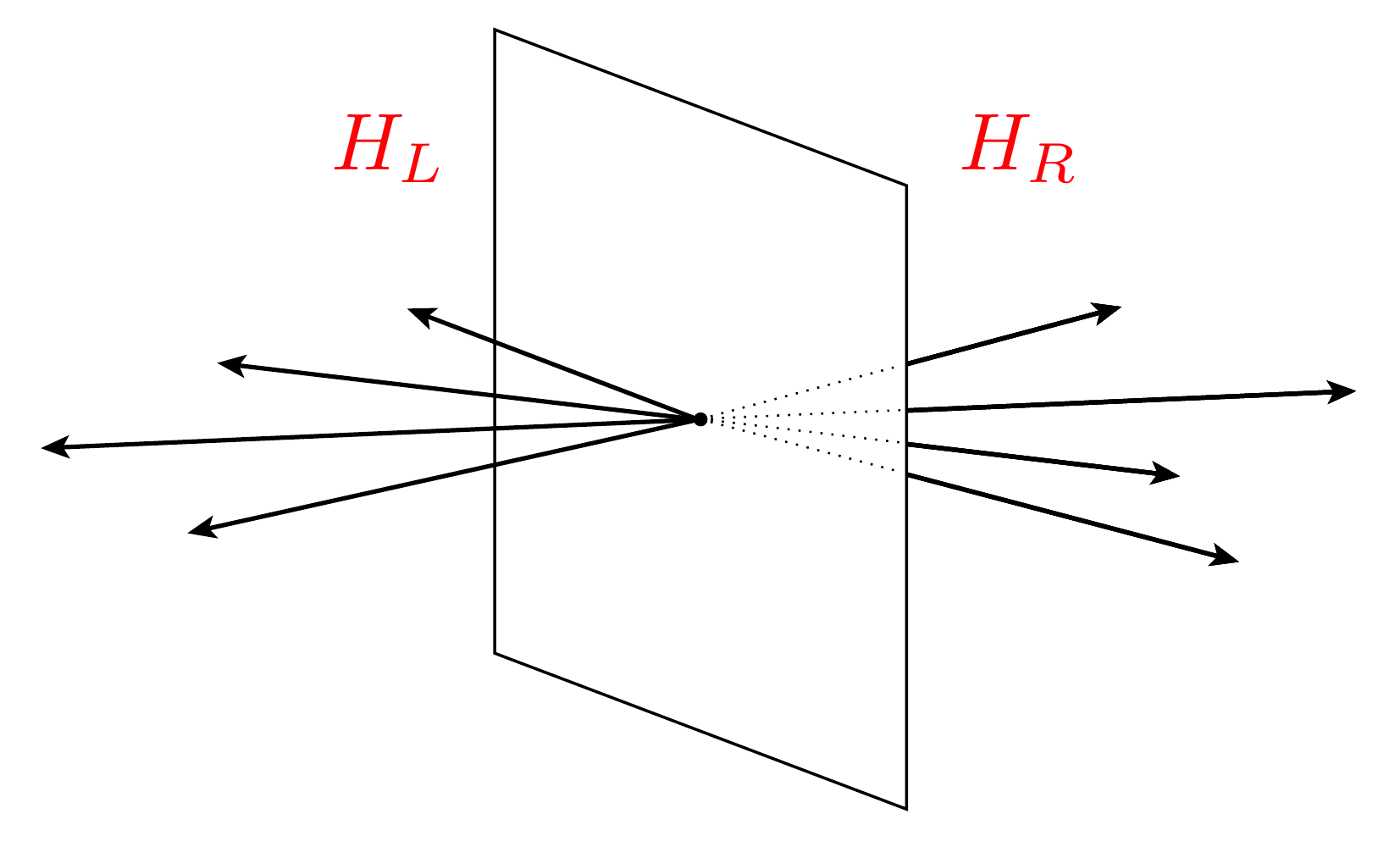}
\includegraphics[width=\columnwidth]{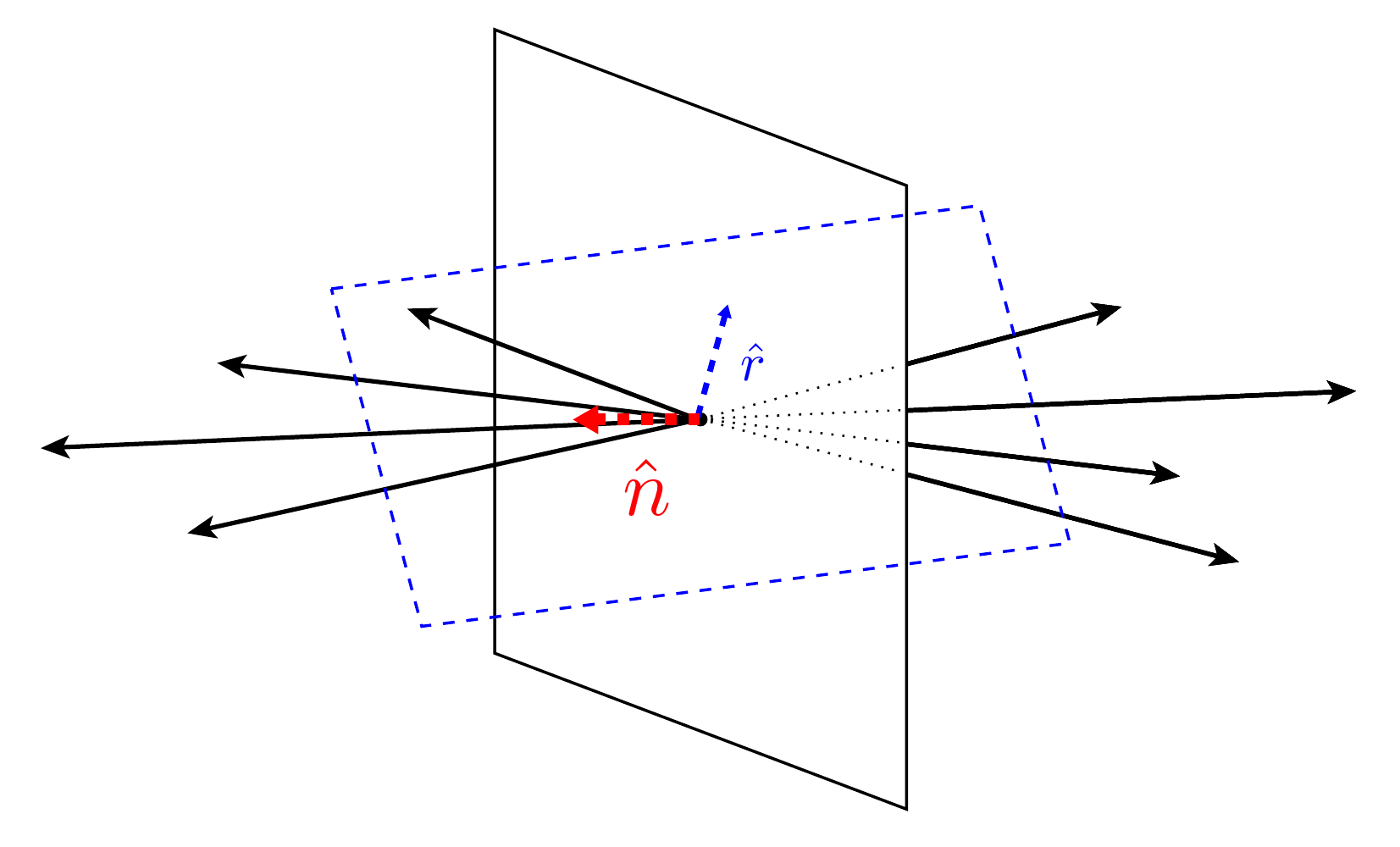}
\caption{Two equivalent definitions of thrust as (left) a partitioning problem and (right) an axis-finding problem.
The best known classical algorithm is based on plane partitioning via a reference axis $\hat{r}$ (which in general differs from the thrust axis).
}
\label{fig:thrust}
\end{figure*}

The core results of this paper are in \Sec{sec:grover}, where we describe quantum algorithms for computing thrust based on Grover search~\cite{Grover:1996:FQM:237814.237866}.
Although naively Grover search offers a square root speedup over any classical search algorithm, in practice Grover search cannot yield sublinear algorithms.
The reason is that data loading over a classical database of size $N$ requires $O(N)$ time, which limits the achievable gains.
That said, if the classical \emph{search space} scales like $O(N^\alpha)$, we can still use the Grover strategy to reduce the \emph{search loop} to $O(N^{\alpha/2})$, though there will be an additional additive (multiplicative) factor of $O(N)$ if data loading has to happen outside (inside) of the loop.
Using the formulation of thrust as a search over reference axes, we show that $\alpha=2$ in the thrust case.
Thus, we can attribute our speedup to the fact that data loading is performed in superposition, which means that it still requires only $O(N)$ time despite working over a search space of size $O(N^2)$.

The precise speed up achievable in our Grover search strategy depends on the assumed quantum computing paradigm.
We implement two models for retrieving and processing the classical data, based on the abstract operations LOOKUP and SUM.
The \emph{sequential computing model} requires $\widetilde{O}(1)$ qubits and results in an $O(N^2)$ thrust algorithm.
Here we use $\widetilde{O}(\cdot)$ to mean that we neglect factors that are polylog in $N$.
The \emph{parallel computing model} requires $\widetilde{O}(N)$ qubits and results in an $O(N\log N)$ thrust algorithm.
Both computing models are applicable to any general problem where the size of the search space scales like $O(N^\alpha)$ with $\alpha\geq 2$, which are precisely the problems that can typically be sped up with a realistic application of Grover search.

In \Sec{sec:advantage}, we assess whether or not there is any quantum advantage for hemisphere jet finding.
Formally, if one has read access to $O(N)$ classical bits but only write access to $O(\log N)$ bits, then one cannot implement the classical sorting strategy in \Sec{sec:classical_sort}.
In that case, there is a quantum advantage for both sequential and parallel computing models.
With write access to $O(N \log N)$ classical bits, though, classical sorting is possible, and the asymptotic performance of our classical and quantum algorithms is identical (up to $\log N$ factors) in both the sequential and parallel cases.
This equivalence appears to be special to algorithms like thrust where the search space scales like $O(N^2)$, and we speculate that larger search spaces might benefit from Grover speedups even if classical sorting is possible.

Finally in \Sec{sec:generalize}, we briefly consider generalizations of our results to jet algorithms more closely related to those used at the \ac{LHC}.
We consider jet function maximization~\cite{Georgi:2014zwa,Ge:2014ova,Bai:2014qca}, showing that, with suitable modifications, it can be written in \ac{QUBO} form for quantum annealing.
We consider stable cone finding in the spirit of \textsc{SISCone}~\cite{Salam:2007xv}, showing how a single-jet variant we dub \textsc{SingleCone} is amenable to quantum search.
We also comment on quantum multi-jet finding motivated by the \textsc{XCone} algorithm~\cite{Stewart:2015waa,Thaler:2015xaa}.
We conclude in \Sec{sec:conclude} with some broader lessons about quantum algorithms for collider physics.

%
%

\section{Definition of Thrust}
\label{sec:def}

We start by defining thrust~\cite{Brandt:1964sa,Farhi:1977sg,DeRujula:1978vmq}, noting that it has multiple equivalent definitions that suggest different algorithmic strategies, as shown in \Fig{fig:thrust}. 
Thrust can be viewed as a partitioning problem, which lends itself naturally to quantum annealing.
Thrust can alternatively be viewed as an axis-finding problem, which we can frame as a quantum search problem.
Both definitions of thrust can be stated in terms of operator norms, and through this lens, they are in fact dual to each other.

\subsection{Thrust as a Partitioning Problem}
\label{subsec:thrustpart}

Consider a set of $N$ three-momenta $\{\vec{p}_i\}$ in their center-of-momentum frame, where $\vec{p}_i =\{p^x_i, p^y_i, p^z_i\}$:
\begin{equation}
\sum_{i = 1}^N \vec{p}_i = 0.
\end{equation}
An intuitive formulation of thrust (though not exactly the original one~\cite{Brandt:1964sa,Farhi:1977sg}) is to separate the particles into a partition $H_L\cup H_R$ such that momenta on each side are as ``pencil-like" as possible.
That is, we seek to maximize the quantity
\begin{equation}
\label{eq:def}
T(H_L) = \frac{2 \left|\sum_{i\in H_L} \vec{p}_i\right|}{\sum_{i = 1}^N |\vec{p}_i|} = \frac{2 \left|\sum_{i\in H_R} \vec{p}_i\right|}{\sum_{i = 1}^N |\vec{p}_i|},
\end{equation}
where the second equality follows from momentum conservation.
The quantity known as ``thrust'' corresponds to the maximum obtainable value:
\begin{equation}
\label{eq:thrust_partition}
T = \max_{H_L} T(H_L).
\end{equation}
The factor of $2$ in \Eq{eq:def} is conventional such that $1/2 \leq T \leq 1$, where $T = 1$ corresponds to a perfectly pencil-like back-to-back configuration and $T = 1/2$ is an isotropic event.

There is an equivalent geometric formulation of \Eq{eq:def} due to \Ref{Brandt1979}.
Consider sequentially summing the three-momenta $\{\vec{p}_i\}$ to form a closed polygon.
Each sequence yields a different polygon, and computing thrust is equivalent to maximizing twice the diagonal of the polygon over all possible polygons, normalized by the circumference of the polygon.
The diagonal splits the polygon into two halves, which yield the partition $H_L\cup H_R$.
The particles in $H_L$ are said to be in the ``left hemisphere jet" and the particles in $H_R$ are said to be in the ``right hemisphere jet''.

This definition immediately suggests a naive, brute-force classical strategy for computing thrust.
We can enumerate all $O(2^N)$ possible partitions (which can be reduced to $O(2^{N-1})$ using momentum conservation), and then we sum the momenta in each to determine the maximum, resulting in an $O(N \, 2^N)$ algorithm.
This is the version of thrust we will use for the quantum annealing formulation in \Sec{sec:qubo}, which corresponds to attacking the problem using quantum brute force.

\subsection{Thrust as an Axis-Finding Problem}
\label{subsec:thrustaxis}

An alternative definition of thrust is as an axis-finding problem, which is a bit closer to the historical definition~\cite{Brandt:1964sa,Farhi:1977sg}.
Let $\hat{n}$ be a unit norm vector and define
\begin{equation}
\label{eq:thrust_axis_finding}
T(\hat{n})=\frac{\sum_{i=1}^N|\hat{n}\cdot\vec{p}_i|}{\sum_{i=1}^N|\vec{p}_i|}.
\end{equation}
Thrust can then be determined by the maximum value of $T(\hat{n})$ over $\hat{n}$:
\begin{equation}
\label{eq:thrust_axis}
T = \max_{|\hat{n}|=1} T(\hat{n}).
\end{equation}
The optimal $\hat{n}$ is known as the \emph{thrust axis}:
\begin{equation}
\hat{n}_\opt \equiv \argmax_{|\hat{n}|=1} T(\hat{n}).
\end{equation}

To gain some intuition for why \Eqs{eq:thrust_partition}{eq:thrust_axis} are equivalent, note that once we find the thrust axis $\hat{n}_{\opt}$, we can partition the particles into those with $\hat{n}_{\opt} \cdot \vec{p}_i > 0$ and those with $\hat{n}_{\opt} \cdot \vec{p}_i < 0$.
(It is an interesting bit of computational geometry to show that $\hat{n}_{\opt} \cdot \vec{p}_i$ can never be exactly zero for a finite number of particles.)  
Said another way, the plane normal to $\hat{n}_{\opt}$ partitions the event into left and right hemispheres.
Starting from a non-hemisphere partition, it is always possible to increase the value of thrust in \Eq{eq:def} by flipping a particle from one side to the other, so the optimal partition will be defined by a plane.
Because of this equivalence between axis finding and plane partitioning, the thrust objective is sometimes written as
\begin{equation}
\label{eq:thrust_def_axis}
T(\hat{n})= \frac{2\sum_{i=1}^N\Theta(\hat{n}\cdot\vec{p}_i)(\hat{n}\cdot\vec{p}_i)}{\sum_{i=1}^N|\vec{p}_i|},
\end{equation}
where the Heaviside theta function picks out particles in just one hemisphere.

Note that the optimal partitioning plane is not unique, since there can be multiple planes that yield the same partition.
We can exploit this fact to find a computationally convenient partitioning plane, defined by a normal \emph{reference axis} $\hat{r}$.
This reference axis will in general be different from the thrust axis $\hat{n}_\opt$ but nevertheless yield the same value of thrust via \Eq{eq:def}.
Specifically, once the optimal partition is known via a reference axis, the thrust axis can be determined from the total three-momentum in the left hemisphere:
\begin{equation}
\label{eq:nopt}
\hat{n}_\opt = \frac{\sum_{i \in H_L} \vec{p}_i}{\left| \sum_{i \in H_L} \vec{p}_i \right|}.
\end{equation}
We will use this reference axis approach for the classical thrust algorithms in \Sec{sec:classical} and for the quantum search strategies in \Sec{sec:grover}.

\subsection{Duality of Thrust Definitions}
\label{subsec:duality}


Using the formalism of {operator norms}, we can show that these two definitions of thrust are in fact dual to each other.

Let $M: \mathbf{V} \rightarrow \mathbf{W}$ be a map from $\mathbf{V}=\mathbf{R}^m$ with norm $\|\cdot\|_\alpha$ to $\mathbf{W}=\mathbf{R}^n$ with norm $\|\cdot\|_\beta$.
The operator norm of $M$, known as the induced $\alpha$-to-$\beta$ norm, is defined as
\begin{equation}
\label{eq:inducednorm}
\|M\|_{\alpha\rightarrow\beta}\equiv\max_{\|v\|_\alpha=1}\|Mv\|_\beta.
\end{equation}
That is, we search over all vectors $v$ in $\mathbf{V}$ with norm 1 and find the maximum
norm for the vector $Mv$ in $\mathbf{W}$.
The case when $\alpha$ and $\beta$ are both the usual $L^2$ norm corresponds to the
largest singular value of $M$, but in general $\|M\|_{\alpha\rightarrow\beta}$ can be
NP-hard to estimate~\cite{BGGLT18}.
By duality, we can rewrite this as
\begin{equation}
\max_{\|v\|_\alpha=1}\|Mv\|_\beta=\max_{\|y\|_{\beta*}=1}\|M^T y\|_\alpha = \|M^T\|_{\beta^*\rightarrow \alpha},
\end{equation}
where $y$ is in $\mathbf{W}^*$, the vector space dual to $\mathbf{W}$, defined as $\mathbf{W}^*=\mathbf{R}^n$ with dual norm $\|\cdot\|_{\beta*}$.
Thus, the $\alpha$-to-$\beta$ norm of $M$ is the same as the $\beta^*$-to-$\alpha$ norm of $M^T$.

In the context of thrust, we are interested in the following norms for a vector $v\in\mathbf{R}^n$:
\begin{align}
\|v\|_1 &=\sum_{i}|v_i|,\\
\|v\|_2 &=\sqrt{\sum_iv_i^2},\\
\|v\|_\infty & =\max_i|v_i|.
\end{align}
These are known, respectively, as the 1-norm, 2-norm, and sup-norm. 
By H\"older's inequality, the space of vectors endowed with the $p$-norm is dual to the space of vectors endowed with the $q$-norm, where $\frac{1}{p} + \frac{1}{q}=1$.
In particular, the 1-norm is dual to the sup-norm, and the 2-norm is dual to itself.

Now consider the matrix $M_{ij}=(\vec{p}_i)_j$, whose rows are the $N$ three-momenta and whose columns are the $p^x$, $p^y$, and $p^z$ components.
This is a map from $\mathbf{R}^3$ to $\mathbf{R}^N$.
Letting $\alpha=2$ and $\beta=1$, the induced 2-to-1 norm of $M$ is
\begin{equation}
\|M\|_{2\rightarrow 1} = \max_{\|\hat{n}\|_2=1}\|M\hat{n}\|_1=\max_{\hat{n}^2=1}\sum_{i=1}^N|\hat{n} \cdot \vec{p}_i|.
\end{equation}
We recognize the last term as the numerator of $T(\hat{n})$ in \Eq{eq:thrust_axis_finding}.
Since the denominator of $T(\hat{n})$ is independent of $\hat{n}$, this is equivalent to the definition of thrust via axis finding in \Sec{subsec:thrustaxis}.
Thus, thrust takes the form of an induced 2-to-1 norm problem.

By duality, with $\beta^*=\infty$, thrust can alternatively be viewed as a sup-to-2 norm problem:
\begin{equation}
\|M^T\|_{\infty\rightarrow 2} = \max_{\|s\|_\infty=1}\|sM\|_2=\max_{s_i\in\{-1, +1\}}\left\Vert\sum_{i=1}^N s_i\vec{p}_i\right\Vert_2.
\end{equation}
This corresponds to the definition of thrust via partitioning in \Sec{subsec:thrustpart}, since setting $s_i=-1$ denotes flipping the orientation of vector $\vec{p}_i$ relative to the partitioning plane, while setting $s_i=1$ retains the orientation of $\vec{p}_i$.

Therefore, we see that the problem of computing thrust in particle physics is in fact a special instance of the more general problem of computing induced matrix norms.
While there exist choices of $\alpha$ and $\beta$ for which efficient algorithms for computing $\|M\|_{\alpha\rightarrow\beta}$ exist for arbitrary $M$, it is believed that the general problem of computing the induced $2$-to-$1$ norm and that of computing the induced $\infty$-to-$2$ norm are both NP-hard~\cite{Ste05, 2010arXiv1001.2613B, 2009arXiv0908.1397H}.
This suggests that thrust is an excellent testbed to explore possible gains from quantum computation.

\subsection{Alternative Duality Derivation}
\label{subsec:alt_duality}

There is alternative language to understand this thrust duality that will be useful for the generalizations in \Sec{sec:generalize}.
This approach is based on \Ref{Thaler:2015uja}, which showed that different jet finding strategies can sometimes be derived from a common meta-optimization problem.

Consider a partition $H$ (not necessarily defined by a plane) with total three-momentum
\begin{equation}
\label{eq:hemisphere_momentum}
\vec{P} = \sum_{i \in H} \vec{p}_i.
\end{equation}
Our analysis is based on the following objective function that depends on both a choice of partition and a choice of axis:
\begin{equation}
\label{eq:meta}
O(\vec{P}, \hat{n})=\hat{n}\cdot\vec{P}+\lambda(\hat{n}^2-1),
\end{equation}
where $\lambda$ is a Lagrange multiplier to enforce that the axis $\hat{n}$ has unit norm.
At this point, $\vec{P}$ and $\hat{n}$ are completely independent entities, and $\hat{n}$ does not play any role in determining the partition $H$.

For fixed $\vec{P}$, we can optimize $O(\vec{P}, \hat{n})$ over $\hat{n}$:
\begin{equation}
\hat{n}_{\opt}=\frac{\vec{P}}{|\vec{P}|}.
\end{equation}
Plugging this into \Eq{eq:meta} yields
\begin{equation}
O(\vec{P}) \equiv O(\vec{P}, \hat{n}_{\opt}) = |\vec{P}|,
\end{equation}
which is (half) of the thrust numerator in \Eq{eq:def}.

For fixed $\hat{n}$, we can optimize $O(\vec{P}, \hat{n})$ over $\vec{P}$ (or equivalently, over the partition $H$):
\begin{equation}
\vec{P}_{\opt} =\sum_{i = 1}^N \Theta(\hat{n}\cdot\vec{p}_i) \, \vec{p}_i.
\end{equation}
Plugging this into \Eq{eq:meta} yields
\begin{equation}
O(\hat{n}) \equiv O(\vec{P}_{\opt}, \hat{n}) =\sum_{i=1}^N \Theta(\hat{n}\cdot\vec{p}_i)(\hat{n}\cdot\vec{p}_i),
\end{equation}
which is (half) of the thrust numerator in \Eq{eq:thrust_def_axis}.

Since the order of optimization is irrelevant to the final optimum, this again shows that the two thrust definitions are dual.
Either way, the maximum value of the objective function will be:
\begin{equation}
O(\vec{P}_{\opt}, \hat{n}_{\opt})=|\vec{P}_{\opt}|,
\end{equation}
which, following \Ref{Brandt1979}, is just the maximum achievable polygon diagonal.

%
%

\section{Classical Algorithms}
\label{sec:classical}

We now describe the best known classical algorithm for thrust in the literature, which requires $O(N^3)$ time, and then show how it can be improved to $O(N^2\log N)$ using a sorting technique from \Ref{Salam:2007xv}.
We start by assuming a sequential classical computing model in this section, and end with a brief discussion of parallel classical computing.

\subsection{Plane Partitioning via a Reference Axis}
\label{sec:classical_paritioning}

The best known classical thrust algorithm~\cite{Yamamoto:1984fd} uses the reference axis approach discussed at the end of \Sec{subsec:thrustaxis}.%
\footnote{Strangely, \Ref{Yamamoto:1984fd} claims $O(N^2)$ usage, which only includes the number of partitions to check, not the computation of thrust itself.}
This is the thrust algorithm implemented in \textsc{Pythia} as of version 8~\cite{Sjostrand:2014zea}.%
\footnote{Version 6 of \textsc{Pythia}~\cite{Sjostrand:2006za} uses a heuristic to approximate thrust, via an iterative procedure that updates the partition starting from seed axes.  While this method converges very quickly, it only finds a local maximum, not the global one~\cite{Brandt1979}, though this may be sufficient for practical applications.  See related discussion in \Ref{Stewart:2015waa}.}
The key realization is that, because of \Eq{eq:nopt}, one only needs to search over inequivalent plane partitions.
Two particles are sufficient to determine a separating plane, so there are $O(N^2)$ inequivalent plane partitions to consider.
For each partition, determining $T(H_L)$ takes $O(N)$, leading to an $O(N^3)$ algorithm.

More specifically, for each pair of particles $\vec{p}_i$ and $\vec{p}_j$, one determines a reference axis $\hat{r}_{ij}$ normal to the plane spanned by them:
\begin{equation}
\hat{r}_{ij} \equiv \frac{\vec{p}_i\times\vec{p}_j}{|\vec{p}_i\times\vec{p}_j|}.
\end{equation}
Then, each particle $\vec{p_k}$ is either assigned to the hemisphere $H_{ij}$ if $\hat{r}_{ij} \cdot \vec{p_k} >0$ or ignored if $\hat{r}_{ij} \cdot \vec{p_k} <0$.
Cases where $\hat{r}_{ij} \cdot \vec{p_k} = 0$ are ambiguous, and we provide a general strategy to deal with this in \Sec{sec:doubling} below.
At minimum, we have to treat the cases where $k = i$ or $j$, which requires testing $2 \times 2 = 4$ possibilities for whether or not $\vec{p}_i$ and/or $\vec{p}_j$ should be included in $H_{ij}$, for a total of $4 N (N -1)$ partitions.
(This can be reduced by a factor of 2 using momentum conservation, since $\hat{r}_{ij}$ and $\hat{r}_{ji}$ define the same hemispheres.)
The final hemisphere jets are determined by the partition that maximizes
\begin{equation}
T_{ij} \equiv T(H_{ij}).
\end{equation}

Note that, in general, none of the $O(N^2)$ reference axes considered will align with the actual thrust axis.
Nevertheless, the \emph{partitions} defined by $\hat{r}_{\opt}$ and $\hat{n}_{\opt}$ will be identical.
(In the idealized case of infinitesimal radiation everywhere in the event, all possible separating planes would be considered, so $\hat{r}_{\opt}$ would then equal $\hat{n}_{\opt}$.)
Once the optimal partition is known, the thrust axis itself is determined by \Eq{eq:nopt}.

In terms of computational complexity, for a fixed hemisphere $H_{ij}$, it takes $O(N)$ time to compute the hemisphere three-momentum in \Eq{eq:hemisphere_momentum}.
The thrust denominator $T_{\rm denom} = \sum_{i=1}^N|\vec{p}_i|$ also takes $O(N)$ time, but it can be precomputed since it is independent of the partition.
Once $\vec{P}_{ij}$ and $T_{\rm denom}$ are known, though, it only takes $O(1)$ time to determine the value of $T_{ij}$:
\begin{equation}
\label{eq:thrust_via_momentum}
T_{ij} = \frac{2 |\vec{P}_{ij}|}{T_{\rm denom}},
\end{equation}
where we used \Eq{eq:nopt} to derive this expression.
For the best known classical algorithm, there are $O(N^2)$ partitions, and we have to do an $O(N)$ computation of $T(H_{ij})$ for each partition, leading to the $O(N^3)$ scaling.
In \Sec{sec:classical_sort}, we can improve on this runtime by iteratively updating $\vec{P}_{ij}$ in a special order.

\subsection{Doubling Trick}
\label{sec:doubling}

To simplify the thrust algorithm, it is convenient to artificially double the number of particles.
Starting from $N$ three-momenta, we create a list of length $2N$ by including both $\vec{p}_k$ and its negative $-\vec{p}_k$.
Because $\vec{p}_k$ and $-\vec{p}_k$ can never be in the same hemisphere, and because of the momentum conservation relation in \Eq{eq:def}, this doubling trick has no effect on the value of thrust.
It does, however, provide us with a convenient way to deal with the four-fold ambiguity above, since we can now define the hemisphere $H_{ij}$ to always include particle $i$ and particle $j$.

To deal with cases where $\hat{r}_{ij} \cdot \vec{p_k} = 0$ (i.e.~any time three or more particles are coplanar), we offset the reference axis by
\begin{equation}
\label{eq:offset}
\hat{r}_{ij} \to \hat{r}_{ij} + \epsilon \vec{q}_{ij}, \qquad  \vec{q}_{ij} \equiv \frac{\vec{p}_i}{|\vec{p}_i|} +  \frac{\vec{p}_j}{|\vec{p}_j|},
\end{equation}
and then take the formal $\epsilon \to 0$ limit.
Specifically, if $\hat{r}_{ij} \cdot \vec{p_k} = 0$, then particle $\vec{p}_k$ is included in $H_{ij}$ if $\vec{q}_{ij} \cdot \vec{p}_k > 0$ and ignored otherwise.

Crucially, \Eq{eq:offset} ensures that $\vec{p}_i$ and $\vec{p}_j$ are always in the hemisphere $H_{ij}$, but $-\vec{p}_i$ and $-\vec{p}_j$ are not.
(One has to be mindful of the pathological situation where $\vec{p}_i$ and $\vec{p}_j$ are exactly anti-parallel, though in this case, thrust is determined by one of the other hemisphere partitions.)
The hemisphere three-momentum is now
\begin{equation}
\label{eq:Pij}
\vec{P}_{ij} = \frac{1}{2}\sum_{k \in H_{ij}} \vec{p}_k,
\end{equation}
where the factor of $\frac{1}{2}$ compensates for the artificial doubling.

We will use this doubling trick repeatedly in this paper, though not for quantum annealing in \Sec{sec:qubo} where it is counter-productive.
To simplify the description of the algorithms, we will leave implicit the treatment of all $\hat{r}_{ij} \cdot \vec{p_k} = 0$ cases via \Eq{eq:offset}.
It is worth mentioning that an alternative way to deal with coplanar configurations is to offset the momenta by a small random amount, but we find the doubling trick to be more convenient in practice since it avoids the four-fold ambiguity automatically.

\subsection{Improvements via Sort}
\label{sec:classical_sort}

The $O(N^3)$ algorithm can be further improved to run in time $O(N^2\log N)$.%
\footnote{We thank Gregory Soyez for discussions related to this point.}
This can be achieved using a strategy from \textsc{SISCone}~\cite{Salam:2007xv} which uses a clever choice of traversal order.
Note that \textsc{SISCone} is intended for proton-proton collisions, whereas our interest here is in electron-positron collisions, but the same basic strategy still applies.

The goal of \textsc{SISCone} is to find conical jet configurations $J$ where the enclosed particles are within a distance $R$ from the cone axis $\hat{n}_J$.
Moreover, these cone jets must be \emph{stable}, meaning that the jet three-momentum $\vec{P}_{J} = \sum_{i \in J} \vec{p}_i$ is aligned with the cone axis $\hat{n}_J$.
Like thrust, \textsc{SISCone} involves solving a partitioning problem where the naive brute-force approach requires $O(N \, 2^N)$ time.
Like for thrust, one can reduce the naive runtime for \textsc{SISCone} to $O(N^3)$ using the fact that two points lying on the circumference of a circle are sufficient to determine the cone constituents.
There is an eight-fold ambiguity in the cone assignments, which we discuss further in \Sec{subsec:siscone}.

The key insight of \Ref{Salam:2007xv} is that one need not recompute $\vec{P}_{J}$ for all $O(N^2)$ candidate cones.
Ignoring the eight-fold ambiguity, let the candidate cones be labeled by $i$ and $j$.
For fixed $i$, one can define a special traversal order for $j$ such that only one particle enters or leaves the cone at a time.
There are $N$ particles labeled by $i$, and for fixed $i$, sorting over $j$ takes $O(N \log N)$ time.
After the initial $O(N)$ determination of $\vec{P}_{J}$ for the first $j$ values in the sorted list, updating the value of $\vec{P}_{J}$ for each $j$ iteration only requires $O(1)$ time, since you need only add the momentum of a point entering the cone or subtract the momentum of a point leaving the cone.
Thus, the final algorithm is $O(N^2 \log N)$.

\begin{figure}
\includegraphics[width=\columnwidth]{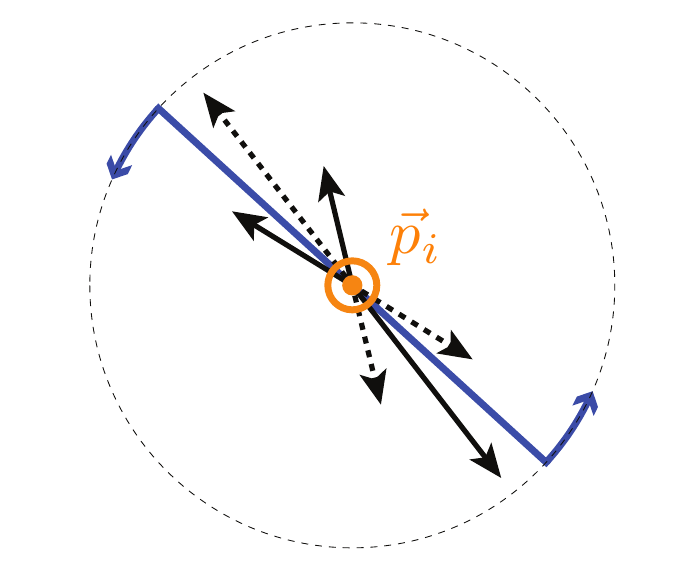}
\caption{Illustration of the sorting algorithm around the $\vec{p}_i$ axis (seen from the top down).  The dashed vectors correspond to the doubling trick.  As the blue partitioning plane sweeps in azimuth, the hemisphere momentum is updated according to \Eq{eq:sweep}.}
\label{fig:sorting}
\end{figure}

We can apply exactly the same sorting strategy to the computation of thrust, as shown in \Fig{fig:sorting}.
The reason is that the reference axis $\hat r_{ij}$ depends only on the cross product $\vec p_i \times \vec p_j$.
This means that for fixed $\vec p_i$, we can choose an ordering of the $\vec p_j$ such that the partitions induced by $\{\hat r_{i1},\ldots,\hat r_{iN}\}$ are specified by a single sortable parameter.
To see this, it is convenient to transform to a coordinate system where $\vec p_i$ points in the $z$ direction, i.e.\ $\vec p_i = |\vec{p_i}| \, (0,0,1)$.
For any $j\neq i$, we can write $\vec p_j$ in spherical coordinates as $\vec p_j = |\vec p_j| \, (\sin \theta_j  \cos \phi_j, \sin \theta_j  \sin \phi_j, \cos \theta_j )$, where $\theta_j$ is the polar angle and $\phi_j$ is the azimuthal angle.
Then $\hat r_{ij} = (-\sin \phi_j, \cos \phi_j ,0)$, so the partition is indeed determined by the single parameter $\phi_j$, independent of $\theta_j$.
Specifically, particle $k$ is in hemisphere $H_{ij}$ if
\begin{equation}
\label{eq:quick_partitioning}
\hat r_{ij}\cdot \hat p_k = \sin \theta_k \, \sin(\phi_k-\phi_j)
\end{equation}
is positive.
This implies that $0 < \phi_k-\phi_j < \pi$, where azimuthal angle differences are calculated modulo $2\pi$.

Furthermore, because of the doubling trick, there is a simple way to determine which particles are in the partition.
With the doubling, there are $2N$ possible choices for $i$, and by \Eq{eq:offset} we know that the doubler $-\vec{p}_i$ cannot be in the same partition as $\vec{p}_i$.
Using the above coordinate system, we can sort the remaining $2N-2$ vectors according to their $\phi$ coordinates, i.e.\ so that $0\leq \phi_{j_1} \leq \phi_{j_2} \leq \cdots \leq \phi_{j_{2N-2}} < 2\pi$.
(In cases where two particles happen to have identical values of $\phi_{j}$, their relative ordering does not matter for the argument below, as long as the doublers are also put in the same order.)
Crucially, for a particle at position $a$ in this sorted list, its doubler (which is $\pi$ away in azimuth) must be at position $a+N-1$.
To see why, note that a hemisphere either contains a particle or contains its doubler, so there must be exactly $N$ particles in each hemisphere.
Particle $i$ is already accounted for, meaning that any candidate partition must contain $N-1$ entries from the sorted list.
Since the sorted list is ordered by azimuth, and since the partitioning is determined by azimuth alone via \Eq{eq:quick_partitioning}, the $N-1$ elements from position $a$ to position $a+N-2$ inclusive must be in a common partition, and the doubler must be the next one on the list.
Therefore, candidate thrust partitions always take the form:
\begin{equation}
\label{eq:Hija_partition}
H_{i,j_a} = \{i,j_a,j_{a+1},\ldots,j_{a+N-2}\}.
\end{equation}
(Note that, as in \Eq{eq:offset}, both particle $i$ and particle $j_a$ are always contained in $H_{i,j_a}$.)

These observations allow us to construct an $O(N^2 \log N)$ algorithm for thrust.
The outer loop involves iterating over all $2N$ choices for $i$.
The inner loop involves the following $O(N\log N)$ algorithm.
We perform the sorting procedure above for fixed $i$, which takes $O(N\log N)$ time.
For the first element in the sorted list, we determine the partition $H_{i,j_1}$ using \Eq{eq:Hija_partition} with $a =1$.
We can readily compute $\vec P_{i,j_1}$ via \Eq{eq:Pij} in time $O(N)$, and then compute the associated thrust value via \Eq{eq:thrust_via_momentum} in $O(1)$.
For the subsequent $2N-3$ elements of the sorted list, we step through them one by one, updating the partition from $H_{i,j_a} = \{i,j_a,j_{a+1},\ldots,j_{a+N-2}\}$ to $H_{i,j_{a+1}} = \{i,j_{a+1},j_{a+2},\ldots,j_{a+N-1}\}$.
In doing so, we need to subtract $\vec{p}_{j_a}$ and add $\vec{p}_{j_{a+N-1}}$ (which is the same as $-\vec{p}_{j_a}$ by the doubling trick), leading to the update step:
\begin{equation}
\label{eq:sweep}
\vec{P}_{i,j_{a+1}} = \vec{P}_{i,j_{a}} - \vec{p}_{j_a},
\end{equation}
where one has to remember the factor of $\frac{1}{2}$ in \Eq{eq:Pij}. 
From the updated momentum, we recompute the associated thrust value via \Eq{eq:thrust_via_momentum} in $O(1)$ time.
The total time from stepping through the $2N-3$ partition momenta is $O(N)$, so the inner loop is dominated just by the initial $O(N\log N)$ sorting step.
The maximum $T_{ij}$ over all $i$ and sorted $j$ determines the final hemisphere jets. 
%
%

\subsection{Parallel Classical Algorithm}
\label{subsec:parallel_classical}

The sorting algorithm above requires $O(N^2\log N)$ operations.
In a model with a single CPU and random-access memory, this corresponds to
time $O(N^2\log N)$ as well.
We can also consider parallel computing models in which the $N$ words of
memory are accompanied by $N$ parallel processors; see \Ref{Vishkin10} for more
discussion of these models.
In this case, we will see that a runtime of $O(N\log N)$ can be achieved.
For simplicity, we do not consider the general case in which the number of
parallel CPUs and the amount of memory  can be varied independently, nor will we
discuss the varying models of parallel computing in \Ref{Vishkin10}.

We briefly sketch here how  the sorting strategy in \Sec{sec:classical_sort} can be
sped up with parallel processors.
There are three main computational bottlenecks:  iterating over all particles $i$ ($C_{\rm iter}$), sorting over particles $j$ for fixed $i$ ($C_{\rm sort}$), and determining the hemisphere constituents over each $j$ for fixed $i$ ($C_{\rm hemi}$), leading to a runtime of $O\big(C_{\rm iter} (C_{\rm sort} + C_{\rm hemi}) \big)$.
For sequential classical computing, we found $C_{\rm iter} = O(N)$, $C_{\rm sort} = O(N \log N)$, and $C_{\rm hemi} = O(N)$.
A parallel computer cannot improve on $C_{\rm iter}$, but there are parallel computing algorithms for sorting~\cite{powers91} and partial sums~\cite{ladner80} that would allow us to achieve $C_{\rm sort} = C_{\rm hemi} = O(\log N)$, leading to a $O(N \log N)$ runtime.
We will compare the quantum and classical parallel architectures in \Sec{sec:advantage}.

%
%

\section{Thrust via Quantum Annealing}
\label{sec:qubo}

The first quantum algorithm we describe is based on quantum annealing~\cite{PhysRevE.58.5355, 2000quant.ph..1106F}.
In a quantum annealer such as the D-Wave system~\cite{dwave}, the solution to an optimization problem is encoded in the ground state of a target Hamiltonian.
Such a Hamiltonian takes the form of an Ising model:
\begin{equation}
H(\{s_i\})=\sum_{i=1}^N h_i \, s_i + \sum_{i<j=1}^N J_{ij} \, s_i \, s_j,
\end{equation}
where each of the $N$ Ising spins $s_i\in\{-1, +1\}$ corresponds to a qubit, and the $\{h_i\}$ and $\{J_{ij}\}$ correspond to programmable weights and couplings between qubits, respectively.

Equivalently, under the transformation $s_i=2x_i-1$, we can frame the optimization problem as a \ac{QUBO} problem, where the objective function takes the form
\begin{equation}
\label{eq:qubo}
O(\{x_i\})=\sum_{i,j=1}^NQ_{ij} \, x_i \, x_j, 
\end{equation}
for $x_i\in\{0,1\}$.
Note that the fact that $i, j$ are now summed with repeated indices and the fact that $x_i^2=x_i$ allow us to absorb the linear terms into the quadratic terms.

For the thrust problem, it is convenient to first define the three-momentum of a candidate partition as
\begin{equation}
\vec{P}(\{x_i\}) = \sum_{i = 1}^N \vec{p}_i \, x_i,
\end{equation}
where $x_i=1$ if particle $\vec{p}_i$ is in the partition and $x_i = 0$ otherwise.
Following \Eq{eq:thrust_via_momentum}, the thrust of this partition is given by
\begin{equation}
T(\{x_i\}) = \frac{2 |\vec{P}|}{T_{\rm denom}} = \frac{2}{T_{\rm denom}} \sqrt{\sum_{i,j = 1}^N \vec{p}_i \cdot \vec{p}_j \, x_i  \, x_j}.
\end{equation}
Because of the square root factor, this is not a QUBO problem, but since the optimal partition is the same for any monotonic rescaling of $T(\{x_i\})$, we can optimize the squared relation:
\begin{equation}
T(\{x_i\})^2 = \frac{4}{T^2_{\rm denom}} \sum_{i,j = 1}^N \vec{p}_i \cdot \vec{p}_j \, x_i  \, x_j,
\end{equation}
which now takes the form of the \ac{QUBO} problem in \Eq{eq:qubo}, as desired.
Finding the ground state of $- T(\{x_i\})^2$ (note the minus sign) is the same as determining thrust.

The space usage of a quantum annealing algorithm is $O(N)$, corresponding to one qubit for each $x_i$.
The annealing time required depends on the spectral gap of the particular Hamiltonian, and we leave the question of determining the spectral gap of the thrust objective function to future work.

%
%

\section{Thrust via Quantum Search}
\label{sec:grover}

We now describe a quantum algorithm for thrust based on Grover search.
We first describe the algorithm in terms of two abstract operations, LOOKUP and SUM, both of which perform data loading in superposition.
Then, we describe two computing models for loading the classical data into quantum memory: the sequential model and the parallel model.
Key to the algorithmic speedups we achieve is the fact that even if quantum data loading takes time $O(N)$, other calculations inside the Grover search loop also take $O(N)$ in both the
classical and quantum models, so we gain from decreasing the effective search space from $O(N^2)$ to $O(N)$.
The sequential model results in an algorithm that requires $O(N^2)$ time and $O(\log N)$ qubits.
The parallel model requires $O(N \log N)$ time and $O(N\log N)$ qubits.
We also assess how the resource requirements of these algorithms scale with the precision of the computation.

\subsection{Algorithm Overview}
\label{subsec:groveroverview}

Our quantum thrust algorithm is based on the quantum maximum finding algorithm
of D\"urr and H\o{}yer~\cite{Durr:1996nx}, which returns the maximum element of an unsorted array with $K$
elements in $O(\sqrt{K})$ time, assuming quantum query access to the array.
This algorithm is itself a generalization of Grover search~\cite{Grover:1996:FQM:237814.237866}.

In this context, quantum query access means that for an array
$A[1],\ldots,A[K]$, we can efficiently perform a unitary operation $U$ such that
\ba U_A \ket{i}\ket{0} = \ket{i} \ket{A[i]},\label{eq:query-access}\ea along
with its inverse $U^\dag$.  The first register, containing $\ket{i}$, should
have dimension at least $K$, so that $\ket{1},\ldots,\ket{K}$ are each
orthogonal states of the register, and the second register should be large
enough to store the values $A[i]$.  Note that \Eq{eq:query-access} does not
fully specify the unitary $U_A$ since it does not specify its action when the
second register is not initially in the state $\ket 0$. One possible way to
define $U_A$ fully is to have $U\ket{i}\ket{x} = \ket{i}\ket{x + A[i]}$ with
addition defined over an appropriately sized finite ring such as
$\mathbb{Z}_2^n$, but this is not necessary for applications such as
in \Refs{Grover:1996:FQM:237814.237866,Durr:1996nx}.  Quantum query access to an
array $A$ is more demanding than simply having $A$ stored on disk, as we will
discuss below.

Recall that Grover search finds one marked item out of an array of $K$ items, assuming the ability to reflect about the marked item.
\Ref{Boyer:1996zf} further extends~\Ref{Grover:1996:FQM:237814.237866} to find one marked item when there are $t > 1$ marked items, assuming the ability to reflect about the multiple marked items.
Generic Grover search then consists of the following steps:
\begin{enumerate}
    \item Prepare the initial state $\ket{\psi_0}=\frac{1}{\sqrt{K}}\sum_{i=1}^K \ket{i}$.
    \item Repeat $O(\sqrt{K/t})$ times:
    \begin{enumerate}
        \item Reflect about the marked states;
        \item Reflect about the initial state $\ket{\psi_0}$.
    \end{enumerate}
\end{enumerate}

\begin{figure*}[t]
\fbox{
\begin{minipage}{6.4in}
\begin{enumerate}
    \item Randomly pick an index $j$ and set $\texttt{curr\_\,max} = j$.
    \item Set $\texttt{iter\_\,count} = 0$ and $\texttt{max\_\,it} = 1$.
    \item While $\texttt{iter\_\,count}<O(\sqrt{K})$:
    \begin{enumerate}
   	\item While $\texttt{max\_\,it}<O(\sqrt{K})$:
   	 \begin{enumerate}
        		\item Prepare the initial state $\ket{\psi_0}= \frac{1}{\sqrt{K}} \sum_{i=1}^K\ket{i}$.
        		\item Choose $\texttt{grov\_\,steps}$ uniformly at random from $\{0,1,...,\texttt{max\_\,it}-1\}$.
        		\item Set $\texttt{iter\_\,count}  = \texttt{iter\_\,count} + \texttt{grov\_\,steps}$.
        		\item Repeat $\texttt{grov\_\,step}$ times:
        		\begin{enumerate}
        			\item Reflect about states satisfying $A[i]>A[\texttt{curr\_\,max}]$;
        			\item Reflect about the initial state $\ket{\psi_0}$.
        		\end{enumerate}
        		\item Measure the first register to obtain index $j$; if $A[j]>A[\texttt{curr\_\,max}]$, set $\texttt{curr\_\,max}=j$ and break.
    	\end{enumerate}
	\item Let $\texttt{max\_\,it} = \mu  \times \texttt{max\_\,it}$, where $\mu$ is a constant between 1 and $4/3$.
    \end{enumerate}
    \item Measure the first register to obtain index $j$; if $A[j]>A[\texttt{curr\_\,max}]$, set $\texttt{curr\_\,max}=j$.
\end{enumerate}
\end{minipage}
}
\caption{Quantum search algorithm due to D\"urr-H\o{}yer to find the index corresponding to the maximum entry of an array $A[i]$ with $K$ elements~\cite{Durr:1996nx}.
The number of Grover steps is chosen at random, since this is a search over an unknown number of marked times~\cite{Boyer:1996zf}.
}
\label{fig:maxfindingalg}
\end{figure*}

When the number $t$ of marked items is unknown, \Ref{Boyer:1996zf} employs an exponential searching algorithm that guesses the number of marked items, increasing the guess by a constant factor each time.
This is a probabilistic algorithm that performs a measurement for each guess, finding a solution in overall expected time $O(\sqrt{K/t})$.

The maximum finding algorithm of~\Ref{Durr:1996nx}, summarized in \Fig{fig:maxfindingalg}, is based on this probabilistic exponential searching algorithm.
It keeps track of the current best maximum seen so far and considers marked states to be those that have a larger array entry value than the current maximum.
It employs the Grover-based exponential searching algorithm of \Ref{Boyer:1996zf} for an unknown number of marked states, performing measurements to obtain the maximum with probability at least $1/2$.
If desired, we can improve the success probability to $1-\eta$ with $\eta>0$, at the cost of an extra $O(\log 1/\eta)$ factor, by performing $O(\log 1/\eta)$ rounds of the algorithm.
%

Our quantum thrust algorithms are then a direct application of quantum maximum finding, but now to an array with $K=O(N^2)$ entries corresponding to the choice of separating plane.
To deal with the four-fold ambiguity, we use the doubling trick of \Sec{sec:classical_sort}, including each original vector $\vec{p_k}$ and its negative $-\vec{p_k}$ in the list of three-momenta to obtain a search space of size $K=4N^2$.
Our problem, now, is to find the maximum value of $T_{ij}$ with $i$ and $j$ each ranging over $2N$ possible indices.
This requires us to be able to load the momentum vectors corresponding to each array index, which means that the quantum algorithm must have some means of accessing the classical data.

\subsection{Data Loading Considerations}

We can describe our quantum thrust algorithms in terms of two abstract operations, LOOKUP and SUM.
Their implementation will be described in \Sec{subsec:groverseq} for the sequential model and \Sec{subsec:groverpar} for the parallel model.
Beyond thrust, these operations are quite general in their application to loading classical data into quantum algorithms.

Note that our search space is of size $O(N^2)$, while data loading over $N$ items in a classical database takes time $O(N)$.
Therefore, we can conceptualize our quantum speedup as resulting from being able to
perform data loading over the \emph{superposition} of search space items.
It is important here that the set of search space items is not the same as the set of data
points.  
In general, any application of Grover search over a search space of size $O(N^{\alpha})$ with $\alpha\geq2$ will result in a square root speedup, whereas for $\alpha<2$, the cost of the algorithm will be dominated by the $O(N)$ data loading cost.

The LOOKUP operation is queried with one index corresponding to a given particle, returning the momentum corresponding to that index:
\begin{align}
\label{eq:LOOKUP}
U_{\rm LOOKUP} \ket{i}\ket{\vec{0}}&=\ket{i}\ket{\vec{p}_i}.
\end{align}
Note that the second register, initialized as $\ket{\vec{0}}$, has to be large enough to store the three-momenta to the desired (qu)bit accuracy.
To make $U_{\rm LOOKUP}$ unitary, we define $U_{\rm LOOKUP} \ket{i}\ket{\vec q} = \ket i \ket{\vec q + \vec p_i}$ for general vectors $\vec q$, where the addition is done
modulo some value larger than the maximum momentum encountered in the problem.
To deal with pairs of particles, we can call $U_{\rm LOOKUP}$ twice on different registers to map $\ket{i} \ket{j} \ket{\vec{0}} \ket{\vec{0}} \to \ket{i} \ket{j} \ket{\vec{p}_i} \ket{\vec{p}_j}.$
This LOOKUP operation will be used to determine all $O(N^2)$ reference axes $\hat{r}_{ij}$, taking $O(N)$ time in the sequential model and $O(\log N)$ time in the parallel model.

The SUM operation returns the sum over all momenta, possibly with a transformation $f(\vec{p}; c)$ applied to each momentum vector:
\begin{equation}
U_{\rm SUM} \ket{c} \ket{0}= \ket{c} \ket{\Sigma_{k=1}^N f(\vec{p}_k; c)},
\end{equation}
where $c$ represents possible control qubits.
From a given reference axis $\hat{r}_{ij}$, SUM will be used to calculate the value of $T_{ij}$.
It is crucial that calculating $T_{ij}$ for fixed $i$ and $j$ takes the same runtime as LOOKUP, i.e.\ $O(N)$ for sequential and $O(\log N)$ for parallel.
Notably, a wide class of collider observables can be computed in linear runtime~\cite{toappearEFM}, even those that would naively scale like a high polynomial power.

Using LOOKUP and SUM, our quantum thrust algorithm is described in \Fig{fig:groveralg}.
As with standard Grover search, we need to be able to reflect about the initial state and the marked states, namely those whose corresponding values of thrust are larger than the best maximum seen so far.
To identify the marked states, we compute thrust for each choice of separating plane, using LOOKUP and SUM to interface the quantum algorithm with the classical data.
We uncompute intermediate steps of our calculations using standard methods (e.g.
Section 3.2 of \Ref{Nielsen:2011:QCQ:1972505}) to make sure that, after computing $T_{ij}$, the system can be reflected about the initial state.

\begin{figure*}[t]
\fbox{
\begin{minipage}{6.8in}
\begin{enumerate}
    \item Randomly pick indices $m,n$ and set $\texttt{curr\_\,max}=(m, n)$.
    \item Compute $\texttt{p\_\,sum}= \frac{1}{2} \sum_{i=1}^{2N}|\vec{p}_i|$.
    \item Set $\texttt{iter\_\,count}=0$ and $\texttt{max\_\,it}=1$.
    \item While $\texttt{iter\_\,count}<O(N)$:
    \begin{enumerate}
    		\item While $\texttt{max\_\,it}<O(N)$:
		\begin{enumerate}
        			\item Prepare the initial state $\ket{\psi_0}= \frac{1}{2N}\sum_{i,j=1}^{2N}\ket{i}\ket{j} \ket{\vec{0}} \ket{\vec{0}} \ket{\hat{0}} \ket{\vec{0}} \ket{0}$.       			
			\item Choose $\texttt{grov\_\,steps}$ uniformly at random from $\{0,1,...,\texttt{max\_\,it}-1\}$.
        			\item Let $\texttt{iter\_\,count}=\texttt{iter\_\,count}+\texttt{grov\_\,steps}$.
        			\item Repeat $\texttt{grov\_\,steps}$ times:
        			\begin{enumerate}
        				\item Call subroutine $\texttt{COMP\_\,T}$ to compute $T_{ij}$:				\[
				\ket{i}\ket{j}\ket{\vec{0}} \ket{\vec{0}} \ket{\hat{0}} \ket{\vec{0}} \ket{0}\mapsto
				\ket{i}\ket{j}\ket{\vec{0}} \ket{\vec{0}} \ket{\hat{0}} \ket{\vec{0}} \ket{T_{ij}}.
				\]
        \item Reflect about states with $T_{ij}>T_{\texttt{curr\_\,max}}$ with a phase factor:
        \begin{align*}
        \ket{i}\ket{j} \ket{\vec{0}} \ket{\vec{0}} \ket{\hat{0}} \ket{\vec{0}} \ket{T_{ij}} \mapsto (-1)^{\Theta(T_{ij} - T_\texttt{curr\_\,max})} \ket{i}\ket{j} \ket{\vec{0}} \ket{\vec{0}} \ket{\hat{0}} \ket{\vec{0}} \ket{T_{ij}} .
        \end{align*}
        \item Uncompute the $T_{ij}$ register to obtain state:  $(-1)^{\Theta(T_{ij} - T_\texttt{curr\_\,max})} \ket{i}\ket{j} \ket{\vec{0}} \ket{\vec{0}} \ket{\hat{0}} \ket{\vec{0}} \ket{0}$.
        \item Reflect about the initial state using $R_0=2\ket{\psi_0}\bra{\psi_0}-I^{\otimes 7}$.
                \end{enumerate}
        \item Measure the $\{i,j, T_{ij}\}$ registers to obtain $\{k, \ell, T_{k\ell}\}$; if $T_{k\ell}>T_\texttt{curr\_\,max}$, set $\texttt{curr\_\,max}=(k,\ell)$ and break.
    	\end{enumerate}
    	\item Let $\texttt{max\_\,it}=\mu  \times \texttt{max\_\,it}$, where $\mu$ is a constant between 1 and $4/3$.
    \end{enumerate}
    \item Measure the $\{i,j, T_{ij}\}$ registers to obtain $\{k, \ell, T_{k\ell}\}$; if $T_{k\ell}>T_\texttt{curr\_\,max}$, set $\texttt{curr\_\,max}=(k,\ell)$.
\end{enumerate}
\end{minipage}
}

\vspace{.1in}

\fbox{
\begin{minipage}{6.8in}
\textbf{Subroutine} $\texttt{COMP\_\,T}$:
\begin{enumerate}
            				\item Load $\vec{p}_i,\vec{p}_j$ using LOOKUP:
            				\[
					\ket{i}\ket{j} \ket{\vec{0}} \ket{\vec{0}} \ket{\hat{0}} \ket{\vec{0}} \ket{0} \mapsto
            				\ket{i}\ket{j}\ket{\vec{p}_i}\ket{\vec{p}_j} \ket{\hat{0}} \ket{\vec{0}} \ket{0}.
            				\]
            				\item Calculate the reference axis via $\hat{r}_{ij}=(\vec{p}_i\times\vec{p}_j)/|\vec{p}_i\times\vec{p}_j|$:
            				\[
           				\ket{i}\ket{j}\ket{\vec{p}_i}\ket{\vec{p}_j}\ket{\hat{0}}\ket{\vec{0}} \ket{0}
				 	\mapsto
            				\ket{i}\ket{j}\ket{\vec{p}_i}\ket{\vec{p}_j}\ket{\hat{r}_{ij}}\ket{\vec{0}} \ket{0}.
           				\]
           				\item Apply SUM, with $f(\vec{p_k}; \hat{r}_{ij})=\{\vec{p_k}/2 \text{ if } \hat{r}_{ij}\cdot
					\vec{p}_k > 0; \vec{0} \text{ if } \hat{r}_{ij}\cdot \vec{p}_k<0 \}$, to obtain hemisphere momentum $\vec{P}_{ij}$:
           		                \[
               				\ket{i}\ket{j}\ket{\vec{p}_i}\ket{\vec{p}_j}\ket{\hat{r}_{ij}}\ket{\vec{0}} \ket{0} \mapsto
				 	\ket{i}\ket{j}\ket{\vec{p}_i}\ket{\vec{p}_j}\ket{\hat{r}_{ij}} \ket{\vec{P}_{ij}} \ket{0}.
               				\]
				         \item Calculate thrust via $T_{ij} = 2 |\vec{P}_{ij}|/\texttt{p\_\,sum}$:
				         \[
				        \ket{i}\ket{j}\ket{\vec{p}_i}\ket{\vec{p}_j}\ket{\hat{r}_{ij}} \ket{\vec{P}_{ij}} \ket{0} \mapsto \ket{i}\ket{j}\ket{\vec{p}_i}\ket{\vec{p}_j}\ket{\hat{r}_{ij}} \ket{\vec{P}_{ij}} \ket{T_{ij}}.
				        \]
            				\item Uncompute registers to obtain state:  $\ket{i}\ket{j} \ket{\vec{0}} \ket{\vec{0}} \ket{\hat{0}} \ket{\vec{0}} \ket{T_{ij}}$.
                          	 \end{enumerate}
	 \end{minipage}
}
	 
\caption{Our Grover-based quantum thrust algorithm, written in terms of the abstract LOOKUP and SUM operations.
The symbols $\ket{\vec{0}}$, $\ket{\hat{0}}$, and $\ket{0}$ refer to initial states for a three-momentum, normalized axis, and real number, respectively.
Note that we have applied the doubling trick from \Sec{sec:doubling}, such that each $\vec{p_k}$ has its negative $-\vec{p_k}$ in the set of three-momenta.
Cases where $\hat{r}_{ij}\cdot \vec{p}_k = 0$ are treated implicitly via \Eq{eq:offset}.
A key difference compared to \Fig{fig:maxfindingalg} is that the quantity to maximize, $T_{ij}$, is calculated quantumly via the $\texttt{COMP\_\,T}$ subroutine.
}
\label{fig:groveralg}
\end{figure*}

Let $C_{\text{LOOKUP}}$ be the asymptotic cost of LOOKUP and $C_{\text{SUM}}$ be the asymptotic cost of SUM.
The runtime of this algorithm is $O\big(N (C_{\text{LOOKUP}}+C_{\text{SUM}}) \big)$ since there is an $O(N)$ outer Grover search loop, while the inner loop is dominated by one application of LOOKUP and one application of SUM.
Note that the computation of the initial guess for the maximum, $T_{mn}$, can be performed in $O(N)$ time classically, while preparation of the initial state and reflection about the initial state can each be performed in $O(\log N)$ time, the time required to perform a Hadamard gate.

\subsection{Sequential Computing Model}
\label{subsec:groverseq}

The first computing model we consider is one in which one gate, classical or quantum, can be executed per time step.
We should think of the classical computer as controlling the overall computation.
In a single time step, it can either (a) perform a classical logic gate, (b) choose a quantum gate or measurement, or (c) read a word from the input (e.g.\ a single momentum).
Another way to think about this model is that we measure cost by the circuit size, i.e.~the total number of gates. 

While fault-tolerant quantum computers are expected to require parallel control
to perform error correction, there are still plausible models in which the cost
of the computation will be proportional to the number of logical gates.  One
possibility is that the cost is dominated by generating magic states or by
long-range interactions.  Another possibility is that we are using a small
quantum computer without fault tolerance, but in an architecture such as a one-dimensional
ion trap, where the available gates are long-range and cannot be
parallelized.

Under this sequential model, the operations LOOKUP and SUM each take $O(N)$ time and require $O(\log N)$ qubits.
Specifically, LOOKUP requires a register of size $O(\log N)$ to store the query index $i$, along with a register to store the requested three-momentum $\vec{p}_i$.
It operates by performing a sequential scan through all $N$ items in the classical
database to fetch and return $\vec{p}_i$.
More concretely, in $O(1)$ time, we can perform $U_{\text{LOOKUP},i}$, defined by
\begsub{LOOKUPi}
U_{\text{LOOKUP},i} \ket{i}\ket{\vec 0} & = \ket{i}\ket{\vec p_i}, \\
U_{\text{LOOKUP},i} \ket{j}\ket{\vec 0} & = \ket{j}\ket{\vec 0}, & \text{if }j\neq i.
\endsub
Then we implement $U_{\text{LOOKUP}}$ in \Eq{eq:LOOKUP} by performing
$U_{\text{LOOKUP},1}U_{\text{LOOKUP},2}\cdots U_{\text{LOOKUP},N}$ in time $O(N)$.
Similarly, SUM takes time $O(N)$ because it also performs one pass through all $N$ items in the classical database while computing and returning the sum $\sum_{i=1}^N f(\vec{p}_i;c)$.

With this implementation of LOOKUP and SUM, with $C_{\text{LOOKUP}} = C_{\text{SUM}} = O(N)$, the Grover-search based thrust algorithm in \Fig{fig:groveralg} requires $O(N^2)$ time and $O(\log N)$ qubits.

\subsection{Quantum Improvements via Sort?}
\label{subsec:groverseq_sort}

One might wonder whether the runtime of the quantum thrust algorithm could be reduced from $O(N^2)$ to $O(N^{3/2}\log N)$, using the same strategy that we used in \Sec{sec:classical_sort} to reduce the classical thrust algorithm time from $O(N^3)$ to $O(N^2\log N)$.
The answer is yes, in principle, but it would require a computing model beyond the sequential one.

Recall that two points define the partitioning plane, and after selecting the first point, we could sort the second point according to a special traversal order.
This allowed us to avoid the $O(N)$ cost of re-summing the momenta for each candidate plane.
Quantum algorithms require $\Omega(N\log N)$ time for sort~\cite{Hoyer2002}, which means that they cannot be used to speed up this part of the classical algorithm.
In principle, though, we could still obtain a Grover square root speedup when searching over the $O(N)$ candidates for the first point determining the partitioning plane.
Combining the $O(\sqrt{N})$ Grover search over the first point with the $O(N\log N)$ sort over the second point would then yield an $O(N^{3/2}\log N)$ overall algorithm.

The challenge here is that to perform quantum sort, all of the data needs to be stored somehow in quantum memory, which goes beyond the sequential computing model above where only one data point is ever accessed in a given time step.
We leave to future work the design of a quantum computing architecture suitable for loading and sorting data from a classical database.

Assuming that such a sort-friendly architecture exists, one might ask about the origin of the $O(N^2 \log N)$ to $O(N^{3/2}\log N)$ speed up.
Such an improvement is only possible since the strategy in \Sec{sec:classical_sort} converts thrust into a \emph{structured search} problem~\cite{2015arXiv150902374M,2019arXiv190610375M}, which evades the naive bounds on quantum search performance.
Of course, no matter the degree of structure, we can never do better than the $O(N)$ cost to examine each data point once.

\subsection{Parallel Computing Model}
\label{subsec:groverpar}

The parallel computing model reduces the time usage of the sequential model at the expense of additional space usage.\footnote{We thank Iordanis Kerenidis for discussions related to this point.}
Under this model, the operations LOOKUP and SUM each take $O(\log N)$ time but require
$O(N\log N)$ qubits.

An abstract version of this model is the standard quantum circuit model, in
which on $N$ qubits we can perform up to $N/2$ two-qubit gates on as many
disjoint pairs of qubits as we like.
A controlling classical computer with the same parallelism can also be used to
process the measurement outcomes and feed the results back in to the quantum
computer.
To implement this in an actual quantum computer, we would need to assume
long-range connectivity but not all-to-all connectivity.
For example, Brierley~\cite{2015arXiv150704263B} describes how connecting each
qubit to four other qubits is enough to simulate full connectivity with $O(\log
N)$ time overhead.
In what follows, we neglect any $O(\log N)$ or other factors from converting the
abstract circuit model to a concrete architecture.

Parallel data retrieval requires first pre-loading all $N$ database items into
the $O(N)$ qubits.
This can be done in $O(1)$ time, since it requires only parallel copy (or CNOT)
operations from the classical bits onto the qubits.
(Even a cost of $O(N)$ at this stage would not change the asymptotic runtime, so
one could also consider input models in which the data could only be accessed
sequentially, such as tape storage.)
This results in the state
\begin{equation}
\label{eq:parallel_storage}
\ket{1}\ket{\vec{0}}\ket{2}\ket{\vec{0}}...\ket{N}\ket{\vec{0}}\mapsto\ket{1}\ket{\vec{p}_1}\ket{2}\ket{\vec{p}_2}...\ket{N}\ket{\vec{p}_N}.
\end{equation}
Note that this is not the same as qRAM~\cite{PhysRevLett.100.160501}, since we
are loading the classical data into a product state once, and not assuming any
kind of query access to the data.

Now, given our pre-loaded data, we can perform LOOKUP in time $O(\log N)$ by performing binary search on the query index $i$ to locate qubits $\ket{i}\ket{p_i}$.
The binary search can be made unitary using a series of $O(N)$ SWAP gates.
Letting $i=i_1i_{2}...i_M$ in binary, if $i_1=1$ we swap the first $N/2$ $(i, p_i)$ pairs with the last $N/2$ $(i, p_i)$ pairs, if $i_2=1$ we swap the first $N/4$ $(i, p_i)$ pairs with the next $N/4$ $(i, p_i)$ pairs, and so on.
After $O(\log N)$ swaps, we end up with qubits $\ket{i}\ket{\vec{p}_i}$ in the first position.
We can then copy $\ket{\vec{p}_i}$ into a blank register and uncompute the swaps.

Similarly, we can perform SUM in time $O(\log N)$ by combining the entries level by level up a binary search tree indexed by $i$, with $O(N)$ additional registers to store the intermediate steps.
That is, we first add all pairs of entries corresponding to indices $i, i'$ where $i_1=i_1', i_2=i_2',...,i_{M-1}=i_{M-1}'$ and $i_M\neq i_M'$.
Then we have $N/2$ entries indexed by $j=j_1j_2...j_{M-1}$, and again we add all pairs of entries corresponding to indices $j, j'$ where $j_1=j_1', j_2=j_2',...,j_{M-2}=j_{M-2}'$ and $j_{M-1}\neq j_{M-1}'$.
Repeating this process $O(\log N)$ times allows us to sum all the entries in parallel.

Thus, the quantum thrust algorithm for the parallel data loading model, with $C_{\text{LOOKUP}} = C_{\text{SUM}} = O(\log N)$, requires $O(N \log N)$ time and $O(N\log N)$ qubits.

%
%

\subsection{Resource Requirements}
\label{sec:resource}

In the above discussion, we focused on the scaling of our Grover-based quantum thrust algorithm in terms of the number of particles $N$.
Here, we want to provide more information on the practical resource requirements for this algorithm in terms of the required precision of the thrust computation.

Thus far, we have been working with data in the form of three-vectors $\vec{p}_i$, where we assumed that the register holding $\vec{p}_i$ is of constant size.
Just how large is this constant, given that using a finite number of qubits would result in digitization error?
For typical collider physics applications, such as anticipated for a future $e^+e^-$ collider, we would want a dynamic range on momenta from the MeV scale (i.e.~per-mille accuracy on GeV-scale hadrons) to the TeV scale (i.e.~the rough energy scale for CLIC), or around six orders of magnitude.
This means $b=\log_2 10^6 \approx 20$ bits of accuracy.
Since we are keeping track of $d=3$ dimensions, the register holding the $\vec{p}_i$ must be of size $db=3b$.
Thus, the total number of qubits required is $O(\log_2 N + db)$ for the sequential algorithm, and $O(N(\log_2 N + db))$ for the parallel algorithm.

To be more specific, the sequential version of the algorithm in \Fig{fig:groveralg} requires 2 registers with $\log_2 N$ qubits, 4 registers with $d b$ qubits, and 1 register with $b$ qubits, apart from any ancillas used in arithmetic operations, for a total of $2\log_2 N + (4d+1)b$ qubits.
For $N = 128$ particles (after the doubling trick), which is reasonable for most $e^+e^-$ applications, this is around 300 qubits.
Such a device is not far beyond current $\approx 50$-qubit computers, so it is naively plausible that the first quantum computer able to run the sequential quantum thrust algorithm (without error correction) could be ready in time to compute realistic thrust distributions at a future $e^+e^-$ collider.
Of course, this depends on the gate connectivity of such a device as well as the achievable coherence time, and as discussed below, circuit depth may be more constraining than the number of qubits.
For the parallel architecture, we need $N (\log_2 N + d b)$ additional qubits for initial data loading (see \Eq{eq:parallel_storage}), though more qubits would most likely be required to simulate full connectivity and to store intermediate steps of the SUM operation.
This points to an $O(10^4)$ qubit device, which is rather optimistic on the 20 year timescale, though this could be made more realistic by preclustering particles to reduce $N$ or by using a smaller value of $b$.

Next, we consider the number of gates required by the Grover-based thrust algorithm.
We first apply $2\log_2 (2N)$ Hadamard gates to obtain the initial state, a uniform superposition over the indices $i$, $j$.
We then apply $O(N)$ iterations of the Grover operator $G$, where $G$ consists of two reflections: the reflection over all states with a thrust value greater than the current maximum, an operation requiring the subroutine \texttt{COMP\_T}, and the reflection about the initial state.
Note that the reflection about the initial state can be effected with an application of $H^{\otimes 2N}$, followed by a reflection about the all-zeros state, followed by an application of $H^{\otimes 2N}$.
The Hadamards require $4\log_2 2N$ gates total, while the reflection about the all-zeros state can be obtained using a controlled-$Z$ operator controlled on having the state $\ket{0}$ in the first $\log_2 N$ registers, which requires $\log_2 N$ CNOT gates.
Similarly, after performing \texttt{COMP\_T}, we can perform the reflection over all states with a thrust value greater than the current maximum using a controlled-$Z$ operator controlled on the $b$ bits representing the thrust value, an operation requiring $b$ CNOT gates.
Thus, the total gate usage of the algorithm scales like $O(N(\log_2 N+C_{\texttt{COMP\_T}}+b))$, where $C_{\texttt{COMP\_T}}$ is the gate cost of the $\texttt{COMP\_T}$ subroutine.

What is $C_{\texttt{COMP\_T}}$?
To estimate this, we consider the steps in \texttt{COMP\_T} from \Fig{fig:groveralg}, noting that these steps consist of either data loading operations like LOOKUP and SUM, or elementary arithmetic operations like addition, multiplication, and division.

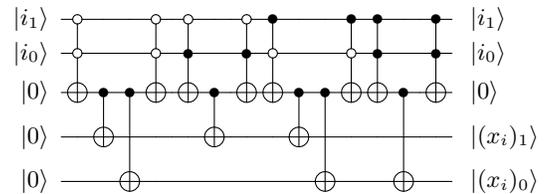
\begin{figure}[t]
\[\Qcircuit @R 1em @C 0.25em {
	&&&&&\lstick{\ket{i_1}} & \ctrlo{1} & \qw & \qw & \ctrlo{1} & \qw & \ctrlo{1} & \qw & \qw & \ctrlo{1} &\ctrl{1} & \qw & \qw & \ctrl{1} & \ctrl{1} & \qw & \qw & \ctrl{1} & \qw & \rstick{\ket{i_1}} \\
	&&&&&\lstick{\ket{i_0}} & \ctrlo{1} & \qw & \qw & \ctrlo{1} &  \qw & \ctrl{1} & \qw & \qw & \ctrl{1} & \ctrlo{1} & \qw & \qw & \ctrlo{1} & \ctrl{1} & \qw & \qw & \ctrl{1} & \qw & \rstick{\ket{i_0}}\\
	&&&&&\lstick{\ket{0}} & \targ & \ctrl{1} & \ctrl{2} & \targ & \qw & \targ & \ctrl{1} & \qw & \targ & \targ & \ctrl{1} & \ctrl{2} & \targ & \targ & \ctrl{2} & \qw & \targ & \qw & \rstick{\ket{0}}\\ 
	&&&&&\lstick{\ket{0}} & \qw & \targ & \qw & \qw & \qw & \qw & \targ & \qw & \qw & \qw & \targ & \qw & \qw & \qw & \qw & \qw & \qw & \qw & \rstick{\ket{(x_i)_1}} \\ 
	&&&&&\lstick{\ket{0}} & \qw & \qw & \targ & \qw & \qw & \qw & \qw & \qw & \qw & \qw & \qw & \targ & \qw & \qw & \targ & \qw & \qw & \qw & \rstick{\ket{(x_i)_0}}\\ 
}\]
\caption{\label{fig:circuit} An example loading circuit mapping $\ket{i}\ket{0}\ket{0}\mapsto \ket{i}\ket{0}\ket{x_i}$.  In our example, $i=i_1i_0$ is two  bits and $(x_0,x_1,x_2,x_3)=(3,2,3,1)$.
The CCNOT gates are drawn with open (closed) circles if they are controlled on the source bit being zero (one).
}
\end{figure}

In step 1, we load $\vec{p_i}$, $\vec{p_j}$ using LOOKUP.
Note that the circuit that implements this looks like the following: first, we have an ancilla bit controlled on each bit in the index register $\ket{i}=\ket{i_{\log_2 N}....i_2i_1i_0}$; that is, we have a $\text{C}^{\log_2 N}\text{NOT}$ gate connecting the ancilla to each index register $\ket{i_k}$.
This requires a total of $\log_2 N$ CNOT gates~\cite{Bar+95}.
Then, controlled on whether or not the ancilla bit is set, we want to transform the blank register $\ket{\vec{0}}$ into the register $\ket{\vec{p_i}}$.
We set each bit of $\vec{p_i}$ controlled on whether or not the ancilla bit is set, so in total we require $db$ CNOT gates.
Finally, we uncompute the ancilla bit by again applying the $\text{C}^{\log_2 N}\text{NOT}$ gate connecting the ancilla to the $\ket{i}$ register, again requiring $\log_2 N$ gates.
In \Fig{fig:circuit}, we give an example circuit for $i=i_1i_0$, indexing two bits corresponding to items 0, 1, 2, 3 with example values.
We have such a circuit for all indices $i$, requiring $O(N(\log_2 N + db))$ gates in total.
For fault-tolerant quantum computers, this procedure can be further
optimized~\cite{LKS18}, but this does not significantly change the
resource scaling.

The remaining steps in \texttt{COMP\_T} involve performing basic arithmetic operations like addition, multiplication, and division.
Circuits for elementary operations like addition and multiplication can be found in~\Ref{Vedral:1995ga}, while fault-tolerant versions can also be found in the literature~\cite{gidney, thapliyal}.
Note that for an input of $n$ bits, addition requires $O(n)$ gates,
while multiplication and division require $O(n^2)$ gates.%
\footnote{Asymptotically faster multiplication circuits exist, but they do not yet
outperform the $O(n^2)$ algorithm until $n \sim 10^{3-4}$; we thank
Craig Gidney for pointing this out.}
Steps 2 and 4 in \texttt{COMP\_T} involve a series of multiplications and divisions with $n=db$ bits, thus requiring $O(d^2b^2)$ gates.
In step 3, we apply SUM controlled on the sign of each $\hat{r}_{ij}\cdot\vec{p_k}$.
Here, we first compute each $\hat{r}_{ij}\cdot\vec{p_k}$ and then set an ancilla bit depending on the sign of the dot product, requiring $O(Nd^2b^2)$ gates total.
Next, for each $\vec{p_k}$ we need to both load the value (using a circuit similar to the one from step 1, requiring $O(N(\log_2 N + db))$ gates total), and then add it to a running sum using an adder circuit if the ancilla bit is set, requiring $O(N(\log_2 N+db))$ gates total.
Thus, $C_{\texttt{COMP\_T}}=O(N(\log_2 N + d^2b^2))$, and the total gate usage of the entire algorithm will scale like $O(N^2(\log_2 N + d^2b^2))$.

Finally we consider circuit depth, which involves considering which gates can be run in parallel.
Note that the $O(N)$ Grover iterations $G$ must come one after the other.
Likewise, within each Grover iteration $G$, the two reflections must come after each other.
The parallelization happens within the subroutine \texttt{COMP\_T}, where we can parallelize LOAD and SUM in the parallel computing model via pre-loading; that is, we execute the loading circuits in parallel so that all the $\vec{p_i}$ are in memory, and then we process the $\vec{p_i}$ in parallel.

First, we perform all $N$ pre-loads in parallel, resulting in a gate depth of $2\log_2 N + db$ gates; this involves performing all $N$ operations in the sequential LOAD operation at once.
After everything has been pre-loaded in parallel memory, we can perform either LOAD or SUM.
To perform LOAD, we want to execute a series of $\log_2 N$ swaps and then a copy, which requires $O(\log_2 N + db)$ CNOT gates, so that the whole LOAD operation has a depth of $O(\log_2 N + db)$.
Meanwhile, to perform the SUM operation after everything has been pre-loaded in parallel memory, we note that we must execute the parallel LOAD operation for each $\vec{p_k}$, then calculate and control on the quantity $\hat{r}_{ij}\cdot\vec{p_k}$ for each $\vec{p_k}$, and then we must finally sum all the $N$ vectors.
The parallel load requires a gate depth of $O(\log_2 N + db)$, while the dot product calculation requires a gate depth of $O(d^2b^2)$.
Finally, we need to perform a series of $\log_2 N$ additions, which requires $db\log_2 N$ gates.
Thus, the SUM operation requires a total circuit depth of $O(db\log_ 2N + d^2b^2)$.
Then $C_{\texttt{COMP\_T}}=O(db\log_2 N + d^2b^2)$, and the circuit depth of the entire parallel algorithm scales like $O(N(db\log_2 N+ d^2b^2))$.
Note that for the sequential model, the circuit depth is just the same as the gate count of $O(N^2(\log_2 N+d^2b^2))$ since we are not running operations in parallel.

Thus, again taking an example with $N=128$ particles (after the doubling trick), we would expect a circuit depth of around $10^7$ gates for the sequential model and $10^5$ for the parallel model.
On a noisy device, we currently do not expect to be able to execute an algorithm requiring more than $10^3$ gates~\cite{nisq}, so again we believe that preclustering particles to reduce $N$ or using a small
value of $b$ could make these algorithms more realistic on a NISQ device.
We note that because circuit depth and qubit usage come at a trade-off, circuit depth is the limiting factor for the sequential model, while qubit usage is the limiting factor for the parallel model.

%
%

\section{Is There a Quantum Advantage?}
\label{sec:advantage}

Starting from the previously best known $O(N^3)$ classical algorithm on a sequential computer, we found an improved $O(N^2 \log N)$ classical algorithm and an $O(N^2)$ quantum algorithm.
Because these scalings are identical up to a $\log N$ factor, one might wonder if there is any real quantum advantage for the task of hemisphere jet finding.

Formally, there is a quantum advantage if we make a rather restricted assumption about the computing model.
The sequential quantum computing model in \Sec{subsec:groverseq} only requires read access to the $O(N)$ classical dataset, whereas the sorting strategy in \Sec{sec:classical_sort} requires write access to $O(N \log N)$ classical bits.
Thus, if one restricts the computing model to have write access to only $O(\log N)$ classical bits, then the classical sorting strategy cannot be implemented.
In that case, the best classical algorithm would be the $O(N^3)$ one from \Ref{Yamamoto:1984fd}, which would be bested by our $O(N^2)$ quantum algorithm.

For any realistic application of thrust, this computing model is overly limited, since data from a single collider event can easily be read into random-access classical memory.
On the other hand, it is not possible to read in the entire LHC dataset into memory, and indeed some collider datasets are only stored on tape drives.
For this reason, there may be interesting quantum advantages for clustering algorithms that act on ensembles of events (instead of on ensembles of particles in a single event).  
See \Ref{Komiske:2019fks} for recent developments along these lines.

For the parallel computing models, there is no formal limit with a quantum advantage, since we need $\widetilde{O}(N)$ (qu)bits with read-write access in
both the quantum and classical cases.
Note that the speed up in the classical and quantum cases come from rather different sources.
Classical sorting splits the $O(N^2)$ search space into an $O(N)$ outer loop and an $O(\log N)$ inner loop.
By contrast, the quantum algorithm searches the $O(N^2)$ search space as a whole in $O(\sqrt{N^2})$ runtime.

This last observation suggests that for even larger search spaces, there might be a quantum advantage even if there exist classical sorting strategies.
If classical sorting can only sort $s$ of the search dimensions, then for an $O(N^\alpha)$ search space, the classical runtime would scale proportional to $O(N^{\alpha - s} \log^s N)$.
The quantum runtime would scale proportional to $O(N^{\alpha/2})$, which would be faster than the classical case for $\alpha > 2s$.
This might be relevant for the $M$-jet finding problem mentioned in \Sec{subsec:xcone} with an $O(N^{2M})$ search space.

%
%

\section{Generalizations}
\label{sec:generalize}

\begin{figure}
\includegraphics[width=\columnwidth]{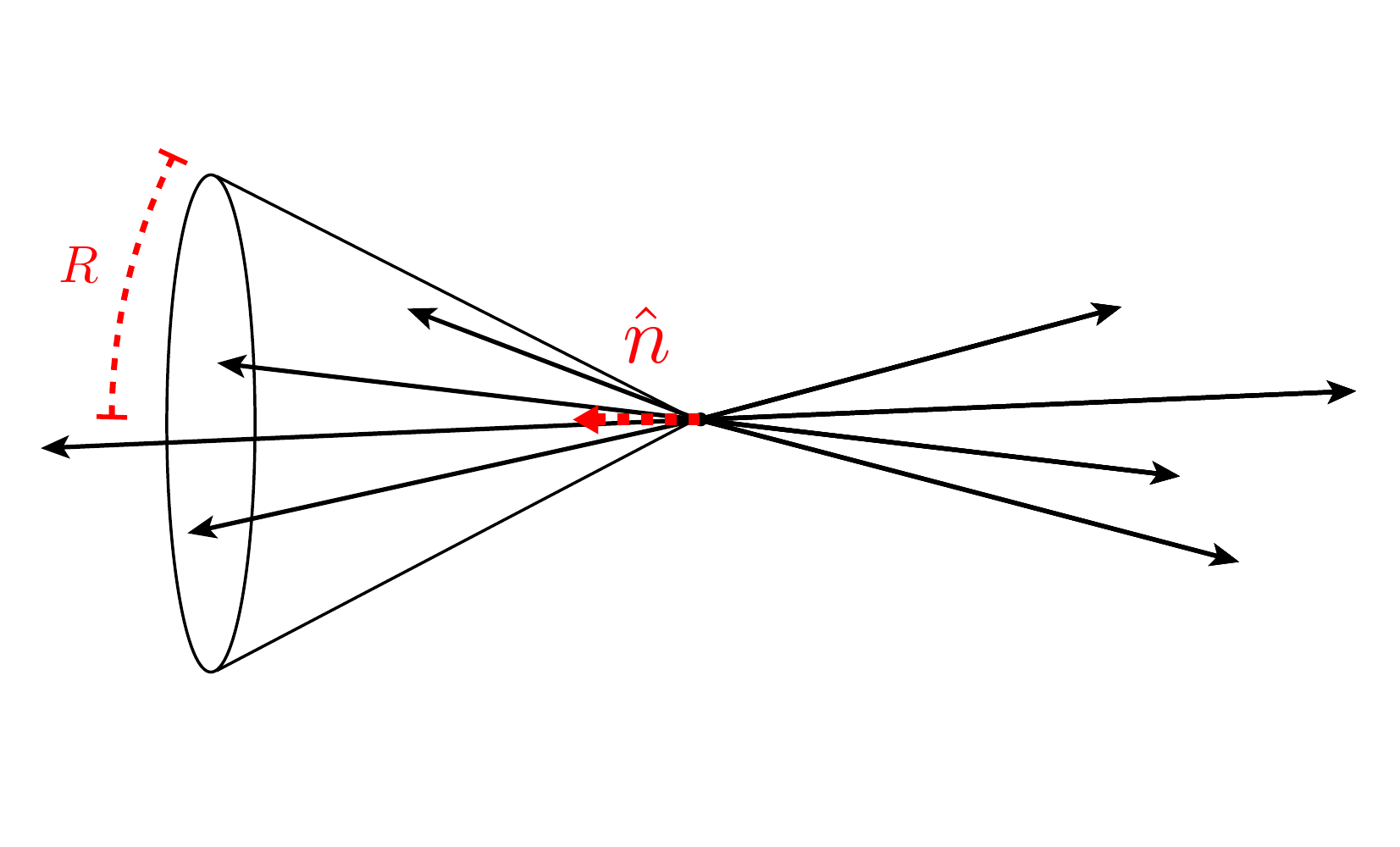}
\caption{Partitioning an event into a stable cone jet of radius $R$ and an unclustered region.  This is the same as \Fig{fig:thrust} when $R = \pi/2$.}
\label{fig:cone}
\end{figure}

In this section, we discuss how to apply the quantum algorithms from \Secs{sec:qubo}{sec:grover} to jet identification methods that generalize thrust.
These algorithms are more closely related to the ones used at the \ac{LHC}, since they involve a jet radius parameter $R$.

We start with algorithms that divide the event into a single jet and an unclustered region, as in \Fig{fig:cone}.
(For the thrust problem, $R = \pi/2$ and the unclustered region is the opposite hemisphere.)
We then mention strategies to identify multiple jets.
To simplify the discussion, we continue to use the $(p^x,p^y,p^z)$ coordinate system for electron-positron collisions, noting that the methods below can be adapted to the standard proton-proton coordinate system of transverse momentum ($p_T$), rapidity ($y$), and azimuth ($\phi$).

\subsection{SingleCone}
\label{subsec:separated_at_birth}

The generalizations we consider are all based on or inspired by the analysis of \Ref{Thaler:2015uja}, which showed that the thrust duality in \Sec{subsec:alt_duality} holds for a one-parameter family of jet finding algorithms.
No matter which dual formulation is used, we refer to this jet finding strategy as \textsc{SingleCone}, since it finds a single stable cone jet of radius $R$.

To match the literature, we use four-vector notation in this section.
The four-momentum of a particle is
\begin{equation}
p^\mu_i = \big(E_i, \vec{p}_i\big),
\end{equation}
where the energy $E_i \equiv p_i^0 = \sqrt{\vec{p}_i^{\,2} + m_i^2}$ depends on the mass $m_i$ of particle $i$.
The four-momentum of a candidate partition $H$ is
\begin{equation}
P^\mu = \sum_{i \in H} p^\mu_i \equiv \big(E, \vec{P}\big),
\end{equation}
where $E \equiv P^0$ is the total energy of the partition.
A light-like axis is given by
\begin{equation}
n^\mu = \big(1,\hat{n} \big),
\end{equation}
with $\hat{n}^2 = 1$.
We contract indices with the mostly minus metric:
\begin{equation}
p_\mu q^\mu = p^0 q^0 - \vec{p} \cdot \vec{q}.
\end{equation}

The \textsc{SingleCone} jet finder is based on maximizing the following objective function~\cite{Thaler:2015uja}:
\begin{align}
\label{eq:meta_R}
O(P^{\mu}, n^{\mu}) &= E - \frac{n_\mu P^\mu}{1 - \cos R}  +\lambda (n_\mu n^\mu),\\
& = \frac{\hat{n} \cdot \vec{P} - E\cos R}{1 - \cos R} +\lambda(\hat{n}^2-1), \nonumber
\end{align}
where $\lambda$ is again a Lagrange multiplier, and we maximize over both the choice of partition and the choice of axis.
The second line makes it clear that $R = \pi/2$ returns the thrust objective function in \Eq{eq:meta}.

Performing the same manipulations as in \Sec{subsec:alt_duality}, the optimum axis (for fixed partition) is
\begin{equation}
\label{eq:stablecone}
n_{\opt}^\mu = \left(1, \frac{\vec{P}}{|\vec{P}|} \right).
\end{equation}
Since the optimum axis is aligned with the jet three-momentum, this is an example of a stable cone algorithm; see \Sec{subsec:siscone} below.
The reduced \textsc{SingleCone} objective function is
\begin{equation}
\label{eq:Georgi}
O(P^{\mu}) \equiv O(P^{\mu}, n_{\opt}^{\mu}) = \frac{|\vec{P}| -E\cos R}{1 - \cos R},
\end{equation}
which is an example of a jet function maximization algorithm~\cite{Georgi:2014zwa, Ge:2014ova, Bai:2014qca}.
The optimum solution partitions the event into a clustered region $H$ and an unclustered region (the complement of $H$).
This definition of the problem naturally lends itself to quantum annealing in \Sec{subsec:georgi}.

Doing the dual manipulation, the optimum partition (for fixed axis) is
\begin{equation}
P_{\opt}^\mu = \sum_{i = 1}^N p^\mu_i \, \Theta(E_i (1 - \cos R) - n_\mu p^\mu_i).
\end{equation}
Writing the Heaviside theta function requirement in three-momentum language,
\begin{equation}
\label{eq:conical}
\frac{\hat{n} \cdot \vec{p}_i}{E_i} > \cos R,
\end{equation}
we see that for massless particles ($E_i = |\vec{p}_i|$), the jet constituents are those within an angular distance $R$ of the jet axis.
For $R = \pi/2$, this yields the thrust hemisphere regions.
The reduced \textsc{SingleCone} objective function is now
\begin{align}
O(n^\mu) &\equiv O(P^{\mu}_{\opt}, n^{\mu})  \nonumber\\
&= \sum_{i = 1}^N E_i - \sum_{i = 1}^N \min \left\{E_i, \frac{n\cdot p_i}{1 - \cos R}\right\}, 
\label{eq:Njetty}
\end{align}
where we dropped the Lagrange multiplier term for compactness.
The second term in \Eq{eq:Njetty} is an example of an $N$-jettiness measure~\cite{Stewart:2010tn,Thaler:2010tr,Thaler:2011gf} with $N = 1$, whose minimum yields the \textsc{XCone} jet algorithm~\cite{Stewart:2015waa,Thaler:2015xaa}.
This definition of the problem naturally lends itself to quantum search in \Sec{subsec:siscone}.

\subsection{Jet Function Maximization}
\label{subsec:georgi}

In the jet function maximization approach of~\Refs{Georgi:2014zwa, Ge:2014ova, Bai:2014qca}, the goal is to optimize $P^{\mu}$ for a global jet function.
The original jet function from~\Ref{Georgi:2014zwa} can be written as
\begin{equation}
\label{eq:origGeorgi}
O_{\rm Georgi}(P_\mu)=E-\frac{1}{2 (1 - \cos R)}\frac{M^2}{E},
\end{equation}
where the jet mass is
\begin{equation}
M^2 \equiv P_\mu P^\mu = E^2 - \vec{P}^2.
\end{equation}
In the limit $M \ll E$, this matches the reduced \textsc{SingleCone} objective function of \Eq{eq:Georgi}, though they yield different optimal jet regions for finite-mass jets.

Since jet function maximization is a kind of partitioning problem, it is natural to try to write these objective functions in QUBO form.
However, the original jet function from \Eq{eq:origGeorgi} is not quadratic since it involves a $1/E$ factor, and the \textsc{SingleCone} function in \Eq{eq:Georgi} is not quadratic since $|\vec{P}|$ involves a square root.
Thus, these cannot be rewritten as QUBO problems without some kind of modification.

In the analysis of \Sec{sec:qubo} for thrust, we got around this issue by squaring the thrust objective function, which nevertheless yielded the same partitioning solution.
This approach does not work in this more general case because of non-quadratic cross terms.

What we can do, however, is square the \textsc{SingleCone} objective in \Eq{eq:Georgi} but only keep the lowest non-trivial term in the $M \ll E$ limit.%
\footnote{We thank Eric Metodiev for discussions related to this point.}
(Squaring and expanding \Eq{eq:origGeorgi} yields the same result.)
This gives the following QUBO objective function:
\begin{align}
\label{eq:QUBOGeorgi}
O_{\rm QUBO}(P_{\mu}) & = E^2-\frac{M^2}{1 - \cos R} \nonumber \\
&= \frac{\vec{P}^2 -E^2\cos R}{1 - \cos R} \nonumber \\
&= \sum_{i,j = 1}^N \left(\frac{\vec{p}_i\cdot\vec{p}_j -E_iE_j\cos R}{1 - \cos R} \right) x_i \, x_j,
\end{align}
where again $x_i\in\{0,1\}$.
Taking $R=\pi/2$ in \Eq{eq:QUBOGeorgi} then recovers the thrust (squared) problem.
It is interesting that \Eq{eq:QUBOGeorgi} has the same form as the generalized jet functions in \Ref{Ge:2014ova} (\Ref{Bai:2014qca}) with $n = 2$ ($\alpha = 2$).

This objective function corresponds to a \ac{QUBO} problem and can thus be solved on a quantum annealer.
It will, however, generally yield a different solution compared to \textsc{SingleCone}.
Unlike \textsc{SingleCone}, which yields perfectly conical jets for massless particles via \Eq{eq:conical}, this QUBO jet finder has an effective jet radius that depends on the mass of the jet~\cite{Ge:2014ova,Bai:2014qca}.
Quadratic objective functions were also explored in \Ref{Bai:2015fka} for jet clustering at the \ac{LHC}.
In future work, we plan to characterize the general phenomenological properties of jets identified using QUBO objectives.

\subsection{Stable Cone Finding}
\label{subsec:siscone}

Stable cone algorithms search over candidate jet regions of radius $R$ and select ones that are ``stable''~\cite{Blazey:2000qt, Ellis:2001aa}, meaning that the center of the jet region aligns with the jet momentum.
As shown in \Eqs{eq:stablecone}{eq:conical}, \textsc{SingleCone} is an example of a stable cone algorithm, which is closely related to \textsc{SISCone}~\cite{Salam:2007xv}.

It is worth emphasizing two key differences between \textsc{SingleCone} and \textsc{SISCone}.
First, \textsc{SingleCone} finds a single jet, whereas \textsc{SISCone} finds all stable cones, and a separate split/merge step is needed to determine the final jet regions.
That said, it is possible to run \textsc{SISCone} in \emph{progressive removal} (PR) mode, where one finds the most energetic stable cone, removes the found jet constituents, and repeats the \textsc{SISCone} procedure on the unclustered particles.
In this way, \textsc{SISCone-PR} acts like an iterated application of \textsc{SingleCone}.
Second, \textsc{SingleCone} finds the jet region with the largest value of \Eq{eq:Georgi} ($= E - O(M^2/E)$), whereas \textsc{SISCone-PR} would typically take the stable cone with the largest plain energy $E$.
As we will see below, though, it is still possible to develop quantum algorithms for stable cones with alternative jet hardness sorting schemes.

It is straightforward to implement the \textsc{SingleCone} algorithm (a.k.a.~\textsc{SISCone-PR} with \Eq{eq:Georgi} ordering) via quantum search.
Just as two points define a partitioning plane, two points are enough to determine a cone region of radius $R$~\cite{Salam:2007xv}.
(This is true up to an eight-fold ambiguity, which is twice that of the thrust case because the two candidate cones are not complements of each other as they are for hemispheres.)
We can use the LOOKUP operation to determine all $O(N^2)$ candidate reference axes (which are not the same as the jet axes, but yield the same partitions).
We can then use SUM to calculate \Eq{eq:Georgi} for a fixed reference axis, since finding $P^\mu$ for the particles in the candidate jet region is a linear operation.
We finally use Grover search to find the partition that maximizes \Eq{eq:Georgi}, and we are guaranteed that the found cone jet will be stable via \Eq{eq:stablecone}. 
This algorithm now has the identical structure to thrust, with the same asymptotic scaling as in \Tab{tab:summary}, taking us from a classical $O(N^3)$ algorithm (without sort) to a quantum $O(N \log N)$ algorithm (with parallel data loading).

Note that the quantum maximum finding algorithm only returns one maximum element of an array, so we cannot use it to speed up an algorithm for identifying all stable cones.
We can, however, use it to find one stable cone with a different objective function from \Eq{eq:Georgi}.
For example, to implement \textsc{SISCone-PR} with standard energy ordering, we can use a subroutine consisting of two SUM operations in series.
The first SUM determines $P^\mu$ for the candidate jet region, while the second SUM finds $\tilde{P}^\mu$ for all particles within a radius $R$ of $P^\mu$.
This subroutine would return $P^0$ if $P^\mu = \tilde{P}^\mu$, while it would return $0$ if $P^\mu \not= \tilde{P}^\mu$.
One would then use Grover search to find the maximum subroutine output, with the same asymptotic quantum gains as in the \textsc{SingleCone} case.

\subsection{Multi-Region Optimization}
\label{subsec:xcone}

Typical collider studies involve more than one jet per event, so it is interesting to ask whether these quantum methods can be adapted to the multi-jet case.
As already mentioned, one can use a PR strategy to identify multiple jet regions, so finding $M$ jets just requires $M$ iterations of the algorithms above.
Except in specialized circumstances, the number of desired jet regions does not grow with $N$ and is at most $O(1/R^2)$, so the runtime of \textsc{SingleCone-PR} would scale linearly with $M$.
That said, we are interested in simultaneously optimizing the jet regions as in \textsc{XCone} algorithm~\cite{Stewart:2015waa,Thaler:2015xaa}, in order to treat the overlapping jet regions in a more sophisticated way than just PR.

The QUBO objective in \Eq{eq:QUBOGeorgi} can be easily generalized to the $M$-jet case using $O\big(N(M+1)\big)$ qubits, suitable for quantum annealing.
Instead of a binary assignment of each particle to the clustered or unclustered region, we can do a one-hot encoding with $M+1$ qubits per particle to indicate their assignment to one of the $M$ jet regions or to the unclustered region.
Specifically, let $x_{ir}\in\{0,1\}$ for $i\in\{1,...,N\}$ and $r\in\{0,1,...,M\}$.
We assign $x_{i0}=1$ if particle $i$ is in the unclustered region, $x_{ir}=1$ if particle $i$ is in jet region $r$ for $r\in\{1,...,M\}$, and $x_{ir}=0$ otherwise.
We then add a penalty term to the objective function such that, for fixed $i$, $x_{ir}=1$ for only one value of $r$.

The multi-jet QUBO objective function is
\begin{align}
O_{\rm QUBO}(\{x_{ij}\}) &= \sum_{r = 1}^M  \sum_{i,j = 1}^N \left(\frac{\vec{p}_i\cdot\vec{p}_j - E_iE_j\cos R}{1 - \cos R} \right) x_{ir} \, x_{jr} \nonumber \\
&\quad + \Lambda^2 \sum_{i=1}^N\left(1-\sum_{r=0}^{M} x_{ir}\right)^2. \label{eq:multijetQUBO}
\end{align}
Here, there is a copy of \Eq{eq:QUBOGeorgi} for each of the $M$ jet regions, taking the schematic form of $O=-\sum_{i,j} Q_{ij} \, x_i \, x_j$.
The coefficient of the penalty term must be taken to be $\Lambda^2 > N \max_{ij} {Q_{ij}}$ to ensure that it is never favorable for a particle to be assigned to more than one jet region.
Because \Eq{eq:multijetQUBO} is quadratic in the momentum, it will not have the same behavior as \textsc{XCone} (which has a linear objective function), though we expect the results to be similar for well-separated jets of comparable energies.
This objective function does not penalize empty jet regions, so it might be interesting to run this algorithm with a large value of $M$ to let the number of non-empty jet regions be determined dynamically.

Compared to the single-jet case, the multi-jet case will likely be more difficult to implement on currently available quantum annealing hardware.
Previous numerical studies~\cite{2018QuIP...17...39K} have shown that clustering problems that use multiple qubits to implement one-hot encoding are prone to errors.
The reason is that on annealing hardware, qubit couplings have a maximum dynamic range, which in turn limits the effectiveness of the $\Lambda$ penalty term.
In practice, this means that annealers often output a fuzzy assignment rather than a hard assignment to one cluster.
We would also like to argue that this problem is conceptual in origin.
The search space of the single-jet QUBO problem is $2^N$, whereas the search space of the multi-jet \ac{QUBO} problem is $2^{MN}$.
However, the \ac{QUBO} quantum search space contains many extra unphysical states, since the actual (non-QUBO) search space is size $M^N=2^{N\log M}$.
While the most natural way to address this would be to use qudits with $d=M$ instead of qubits, such hardware is not currently available.

Turning to the quantum search case, finding $M$ conical jet regions naively requires searching a space of $O(N^{2M})$, with the added complication of needing to treat overlapping jet regions.
We are unaware of any classical approach to this problem apart from brute force, though one expects an $O(N^{2M+1})$ algorithm for the \textsc{XCone} objective should be feasible, though it likely requires a more sophisticated treatment of reference axes.
(The current implementation of \textsc{XCone} in \textsc{FastJet contrib} 1.041~\cite{Cacciari:2011ma,fjcontrib} only finds a local minimum starting from suitable seed axes.)
Using quantum search with sequential (parallel) data loading, one might hope that this could be improved to $O(N^{M+1})$ ($O(N^M \log N)$), though one would have to generalize the LOOKUP and SUM operations to deal with the multi-jet case.
At minimum, LOOKUP would have to load the momenta into $2M$ registers (to label the candidate partitions), and SUM would have to have $M$ distinct outputs (for each of the $M$ jet regions).
Even with quantum gains, this is computationally daunting, motivating future studies of multi-jet algorithm whose computational complexity grows only polynomially with $M$.

%
%

\section{Conclusions}
\label{sec:conclude}

In this work, we demonstrated how quantum computers could be applied to a realistic collider physics problem, which requires interfacing a classical data set with a quantum algorithm.
We focused on maximizing thrust to identify hemisphere jets, but the quantum methods developed here are relevant to a broader range of optimization and cluster-finding problems.
The asymptotic performance of our quantum annealing and quantum search algorithms are summarized in \Tab{tab:summary}.
We found a way to improve the previously best known $O(N^3)$ classical thrust algorithm to an $O(N^2)$ sequential quantum algorithm.
Along the way, we found an improved $O(N^2 \log N)$ classical algorithm, based on the sorting strategy of \Ref{Salam:2007xv}.
Both the quantum and improved classical algorithms can be implemented on parallel computing architectures with asymptotic $O(N \log N)$ runtime.
Formally, we found a quantum advantage, but only when assuming a computing model with read access to $O(N)$ (qu)bits but write access to only $O(\log N)$ (qu)bits.

Going beyond thrust, we briefly generalized our quantum methods to handle structurally similar jet clustering algorithms.
These involve maximizing an objective function with a radius parameter $R$, which partitions the event into a conical jet region and an unclustered region.
While we focused on electron-positron collisions, it is known how to adapt these methods to proton-proton collisions~\cite{Bai:2014qca,Thaler:2015uja,Bai:2015fka}.
In future work, we plan to investigate the phenomenological performance of these ``quantum friendly'' jet algorithms at the \ac{LHC}, to assess whether they offer improved physics performance relative to hierarchical clustering schemes like anti-$k_t$.

The main take home message from this work is that the overhead of data loading must be carefully accounted for when evaluating the potential for quantum speedups on classical datasets.
In  many ways, optimization-based jet algorithms are an ideal platform to think about quantum algorithms for collider physics, since these problems tend to involve searching over a large space of possibilities, $O(N^\alpha)$ with $\alpha\geq 2$, and therefore benefit from Grover search methods.
By contrast, even though the number of events in a collider data sample ($N_{\rm events}$) is usually much larger than the number of final-state particles in a jet, typical collider tasks like filling a histogram involve $O(N_{\rm events})$ operations, such that data loading is already the limiting factor.
On the flip side, this motivates further quantum investigations into classically $O(N^2_{\rm events})$ data manipulation strategies, such as the metric space approach recently proposed in \Ref{Komiske:2019fks}, since they might be reducible to $O(N_{\rm events})$ quantum algorithms under suitable circumstances.
We also note that Grover search is limited to a square-root speedup on unstructured search, whereas collider data has additional structures like symmetries and heuristics which might lead to further quantum gains.

\begin{acknowledgments}
We would like to thank Howard Georgi, Craig Gidney, Iordanis Kerenidis, Patrick Komiske, Andrew Larkoski, Eric Metodiev, Stephen Mrenna, Benjamin Nachman, Gavin Salam, Matthew Schwartz, Gregory Soyez, and Jean-Roch Vlimant for helpful conversations and suggestions.
This work was supported by the Office of High Energy Physics of the U.S. Department of Energy (DOE) under Grant Nos. DE-SC0012567 and DE-SC0019128 (QuantISED).
AYW is additionally supported by a Computational Science Graduate Fellowship (CSGF) from the DOE.
AWH is additionally supported by NSF grants CCF-1452616, CCF-1729369, and PHY-1818914; ARO contract W911NF-17-1-0433; and the MIT-IBM Watson AI Lab.
JT is additionally supported by the Simons Foundation through a Simons Fellowship in Theoretical Physics, and he benefited from the hospitality of the Harvard Center for the Fundamental Laws of Nature and the Fermilab Distinguished Scholars program.

\end{acknowledgments}

\bibliography{quantum_thrust}

\end{document}